\pdfoutput=1
\documentclass[iop]{emulateapj}
\usepackage{apjfonts}
\usepackage{amssymb}
\usepackage{amsmath}
\usepackage{ctable}

\newcommand{\be}{\begin{equation}}
\newcommand{\ee}{\end{equation}}

\newcommand{\msun}{M_{\sun}}

\newcommand{\paperone}{Paper {\small I}}
\newcommand{\papertwo}{Paper {\small II}}

\newcommand{\ffirst}{f_{f}}
\newcommand{\Mach}{\mathcal{M}}
\newcommand{\Machcompressive}{\Mach_{c}}
\newcommand{\cbeta}{\zeta}
\newcommand{\mbeta}{\beta}
\newcommand{\Sprime}{S^{\prime}}
\newcommand{\Pfragint}{P_{\rm frag}^{\rm int}}
\newcommand{\rhocrit}{\rho_{\rm crit}}
\newcommand{\rhodisk}{\rho_{0}}

\newcommand{\appendixcolumns}{\twocolumngrid}

\newcommand\plotonesize[2]
 {\centering \leavevmode \includegraphics[width={#2\columnwidth}]{#1}}
\newcommand{\plotsidesize}[2]
 {\centering \leavevmode \includegraphics[width={#2\textwidth}]{#1}}
\shorttitle{Gravo-Turbulent Fragmentation}
\shortauthors{Hopkins \&\ Christiansen}
\slugcomment{Submitted to ApJ, May, 2013}
\begin{document}
\title{\vspace{-1.0cm}Turbulent Disks are Never Stable: \\
Fragmentation and Turbulence-Promoted Planet Formation}
\author{Philip F.\ Hopkins\altaffilmark{1,2} \&\ Jessie L.~Christiansen\altaffilmark{3}}
%\author[Hopkins]{
%\parbox[t]{\textwidth}{ 
%Philip F.~Hopkins\thanks{E-mail:phopkins@caltech.edu}\altaffilmark{1,2},
%\&\ Jessie L.~Christiansen\altaffilmark{3}} 
%\vspace*{6pt} \\
\altaffiltext{1}{TAPIR, Mailcode 350-17, California Institute of Technology, Pasadena, CA 91125, USA; E-mail:phopkins@caltech.edu} %\\
\altaffiltext{2}{Department of Astronomy and Theoretical Astrophysics Center, University of California Berkeley, Berkeley, CA 94720} %\\
\altaffiltext{3}{SETI Institute/NASA Ames Research Center, M/S 244-30, Moffett Field, CA 94035}
%\vspace{-0.7cm}} 
%}

%\date{Submitted to MNRAS, December, 2012\vspace{-0.6cm}}
%\begin{document}
%\maketitle
%\label{firstpage}

\begin{abstract}
%\vspace{-0.3cm}

A fundamental assumption in our understanding of disks is that when the Toomre parameter $Q\gg1$, the disk is stable against fragmentation into smaller, self-gravitating objects (and so cannot, for example, form planets via direct collapse). However, if disks are turbulent, this criterion neglects a broad spectrum of stochastic density fluctuations that can produce rare but extremely high-density local mass concentrations that will easily collapse. We have recently developed an analytic framework to predict the statistics of these fluctuations. Here, we use these models to consider the rate of fragmentation and mass spectrum of fragments formed in a turbulent, Keplerian disk (e.g.\ a proto-planetary or proto-stellar disk). Turbulent disks are never completely stable: we calculate the (always finite) probability of the formation of self-gravitating structures via stochastic turbulent density fluctuations (compressions, shocks, etc.) in such a disk. Modest sub-sonic turbulence above a minimum Mach number $\Mach\gtrsim0.1$ is sufficient to produce a few stochastic fragmentation or ``direct collapse'' events over $\sim$Myr timescales, even if $Q\gg1$ and cooling is relatively ``slow'' ($t_{\rm cool}\gg t_{\rm orbit}$). In trans-sonic turbulence ($\Mach\sim1$) this extends to $Q$ as large as $\sim100$. We derive the ``true'' $Q$ criterion needed to suppress such events (over some timescale of interest), which scales exponentially with Mach number. We specify this to cases where the turbulence is powered by MRI, convection, or density/spiral waves, and derive equivalent criteria in terms of $Q$ and/or the disk cooling time. In the latter case, cooling times as long as $\gtrsim 50\,\Omega^{-1}$ may be required to completely suppress this channel for collapse. These gravo-turbulent events produce a mass spectrum concentrated near the Toomre mass $\sim (Q\,M_{\rm disk}/M_{\ast})^{2}\,M_{\rm disk}$ (spanning rocky-to-giant planet masses, and increasing with distance from the star), with a wider mass spectrum as $\Mach$ increases.  We apply this to plausible models of proto-planetary disks with self-consistent cooling rates and temperatures, and show that, beyond radii $\sim1-10\,$au, no disk temperature can fully suppress stochastic events. For theoretically expected temperature profiles, even a minimum mass solar nebula could experience stochastic collapse events, provided a source of turbulence with modest sub-sonic Mach numbers. 

\end{abstract}

%\begin{keywords}
\keywords{
protoplanetary discs --- planets and satellites: formation --- accretion, accretion disks --- hydrodynamics --- instabilities --- turbulence --- star formation: general --- galaxies: formation 
%\vspace{-1.0cm}
}
%\end{keywords}

%\vspace{-1.1cm}
\section{Introduction}
\label{sec:intro}

\subsection{The Problem: When Do Disks Fragment?}
\label{sec:intro:problem}

Fragmentation and collapse of self-gravitating gas in turbulent disks is a process central to a wide range of astrophysics, including planet, star, supermassive black hole, and galaxy formation. The specific case of a ``marginally stable,'' modestly turbulent Keplerian disk is particularly important, since this is expected in proto-planetary and proto-stellar disks as well as AGN accretion disks. There has been considerable debate, especially, regarding whether planets could form via ``direct collapse'' -- fragmentation of a self-gravitating region in a proto-planetary disk -- as opposed to accretion onto planetesimals \citep[see e.g.][and references therein]{boley:2006.protoplanetary.disk.w.cooling,boley:2009.twomode.giant.planet.formation,dodson-robinson:2009.giant.planet.dircoll.longorbit,cai:2008.protoplanet.disk.w.rad,cai:2010.disk.instab.rt,vorobyov:2010.dircoll.migration,johnson:2010.constraint.direct.coll.giant.planets,boss:2011.disk.instab.vs.obs,stamatellos:2011.disk.frag.easier,dangelo:2011.giantplanet.form.book,forgan:2013.pop.synth.planet.disc.frag}. Even if this is not a significant channel for planet formation, it is clearly critical to understand the conditions needed to {\em avoid} fragmentation. This is especially demanding because such fragments need only form an order-unity number of times over the millions of dynamical times most proto-planetary disks survive, to account for a significant fraction of planets. 

Yet there is still no consensus in the literature regarding the criteria for ``stability'' versus fragmentation in any disk, especially in nearly-Keplerian proto-planetary disks. The classic Toomre $Q$ criterion:
\be
Q \equiv \frac{\sigma_{\rm g}\,\kappa}{\pi\,G\,\Sigma_{\rm gas}} > 1
\ee
where $\sigma_{\rm g}$ is the gas velocity dispersion, $\kappa\sim\Omega$ the epicyclic frequency, and $\Sigma_{\rm gas}$ the gas surface density \citep{toomre:Q,toomre:spiral.structure.review,goldreich:1965.spiral.stability} appears to offer some guidance. And indeed, for $Q\ll1$, most of the mass in disks fragments into self-gravitating clumps in roughly a single crossing time. But this was derived for a smooth, homogeneous disk, dominated by thermal pressure with no cooling, so does not necessarily imply stability in any turbulent system. \citet{gammie:2001.cooling.in.keplerian.disks} studied a more realistic case of a turbulent disk with some idealized cooling, and showed that if the cooling time exceeded a couple times the dynamical time, the disk could maintain its thermal energy (via dissipation of the turbulent cascade), with a steady-state $Q\gtrsim1$ and transsonic Mach numbers (powered by local spiral density waves), thus avoiding runway catastrophic fragmentation. 

However, many subsequent numerical simulations have shown that, although catastrophic, rapid fragmentation may be avoided when these criteria are met, simulations with larger volumes (at the same resolution) and/or those run for longer timescales still {\em eventually} form self-gravitating fragments, even with cooling times as long as $\sim50$ times the dynamical time \citep[see references above and e.g.][]{rice:2005.disk.frag.firstlook,rice:2012.convergence.disk.frag,meru:2011.frag.crit.planet.disk,meru:2011.nonconvergence.disk.frag.time,meru:2012.nonconvergence.disk.frag,paardekooper:2011.edge.fx.disk.frag,paardekooper:2012.stochastic.disk.frag}. Increasing spatial resolution and better resolution of the turbulent cascade (higher Reynolds numbers) appear to exacerbate this -- leading to the question of whether there is ``true'' stability even at infinitely long cooling times \citep{meru:2012.nonconvergence.disk.frag}. And these simulations are still only typically run for a small fraction of the lifetime of such a disk (or include only a small fraction of the total disk mass, in shearing-sheet models), and can only survey a tiny subsample of the complete set of parameter space describing realistic disks. 

%\vspace{-0.5cm}
\subsection{The Role of Turbulent Density Fluctuations}

Clearly, the theory of disk fragmentation requires revision. But almost all analytic theoretical work has assumed that the media of interest are homogeneous and steady-state \citep[though see][and references above]{kratter:2011.disk.frag.criteria}, despite the fact that perhaps the most important property of turbulent systems is their inhomogeneity. In contrast to the ``classical'' homogenous models, \citet{paardekooper:2012.stochastic.disk.frag} went so far as to suggest that fragmentation when $Q\gtrsim1$ may be a fundamentally stochastic process driven by random turbulent density fluctuations -- and so can never ``converge'' in the formal sense described by the simulations above. That said, simulations of idealized turbulent systems in recent years have led to important breakthroughs; in particular, the realization that compressible, astrophysical turbulence obeys simple inertial-range velocity scalings \citep[e.g.][]{ossenkopf:2002.obs.gmc.turb.struct,federrath:2010.obs.vs.sim.turb.compare,block:2010.lmc.vel.powerspectrum,bournaud:2010.grav.turbulence.lmc}, and that -- at least in isothermal turbulence -- the density distribution driven by stochastic turbulent fluctuations develops a simple shape, with a dispersion that scales in a predictable manner with the compressive Mach number \citep{vazquez-semadeni:1994.turb.density.pdf,padoan:1997.density.pdf,scalo:1998.turb.density.pdf,ostriker:1999.density.pdf}. 

Recently, in a series of papers \citep{hopkins:excursion.ism,hopkins:excursion.clustering,hopkins:excursion.imf,hopkins:excursion.imf.variation,hopkins:frag.theory}, we showed that the excursion-set formalism could be applied to extend these insights from idealized simulations, and analytically calculate the statistics of bound objects formed in the turbulent density field of the ISM. This is a mathematical formulation for random-field statistics (i.e.\ a means to use the power spectra of turbulence to predict the statistical real-space structure of the density field), well known from cosmological applications as a means to calculate halo mass functions and clustering in the ``extended Press-Schechter'' approach from \citet{bond:1991.eps}. This is a well-known theoretical tool in the study of large scale structure and galaxy formation, and underpins much of our analytic understanding of halo mass functions, clustering, mergers, and accretion histories \citep[for a review, see][]{zentner:eps.methodology.review}. The application to turbulent gas therefore represents a means to calculate many interesting quantities analytically that normally would require numerical simulations. In \citet{hopkins:excursion.ism} (hereafter \paperone), we focused on the specific question of giant molecular clouds (GMCs) forming in the ISM, and considered the simple case of isothermal gas with an exactly lognormal density distribution. We used this to predict quantities such as the rate of GMC formation and collapse, their mass function, size-mass relations, and correlation functions/clustering, and showed that these agreed well with observations. In \citet{hopkins:frag.theory} (hereafter \papertwo), we generalized the models to allow arbitrary turbulent power spectra, different degrees of rotational support, non-isothermal gas equations of state, magnetic fields, intermittent turbulence, and non-Gaussian density distributions; we also developed a time-dependent version of the theory, to calculate the rate of collapse of self-gravitating ``fragments.'' 

%\vspace{-0.5cm}
\subsection{Paper Overview}

In this paper, we use the theory developed in \paperone\ \&\ \papertwo\ to calculate the statistics of fragmentation events in Keplerian, sub and trans-sonically turbulent disks, with a particular focus on the question of fragmentation in proto-planetary disks. We develop a fully analytic prediction for the probability, per unit mass and time, of the formation of self-gravitating fragments (of a given mass) in a turbulent disk. In addition to providing critical analytic insights, this formulation allows us to simultaneously consider an enormous dynamic range in spatial, mass, and timescale, and to consider extremely rare fluctuations (e.g.\ fluctuations that might occur only once over millions of disk dynamical times), which is impossible in current numerical simulations.\footnote{{It is important to clarify, when we refer to ``large fluctuations,'' are not referring to extremely large, single-structure ``forcing'' events (e.g.\ a very strong shock on large scales). In fact, we assume that the probability of such ``positive intermittency events'' (isolated, large amplitude Fourier modes) is vanishingly small ($p_{+}=0$, in the language of the intermittency models discussed in \S~\ref{sec:turb.density} and \citealt{hopkins:frag.theory,hopkins:2012.intermittent.turb.density.pdfs}). What we calculate is the (rare) probability of many small, independent fluctuations on different scales acting, by random chance, sufficiently ``in phase'' so as to produce a large density perturbation. If ``positive intermittency events'' do occur, it may significantly increase the probability of rare collapse events.}} We will show that this predicts ``statistical'' instability, fragmentation, and the formation of candidate ``direct collapse'' planets/stars even in the ``classical'' stability regime when $Q\gg1$ and cooling is very slow. However, these events occur stochastically, separated by much larger average timescales than in the catastrophic fragmentation which occurs when $Q < 1$. We will show that this naturally explains the apparently discrepant simulation results above, and may be important over a wide dynamic range of realistic disk properties for formation of both rocky and giant planets.

%\vspace{-0.5cm}
\section{``Classical'' vs.\ ``Statistical'' Stability}

Motivated by the discussion above, and the models developed in this paper, we will introduce a distinction between two types of (in)stability: ``classical'' and ``statistical.'' 

By ``classical'' instability, we refer to the traditional regime where $Q<1$, in which small perturbations to the disk generically grow rapidly. A ``classically unstable'' disk will fragment catastrophically, with a large fraction of the gas mass collapsing into self-gravitating objects on a few dynamical times. A ``classically stable'' disk will survive many dynamical times in quasi-steady state, and gas at the mean density will not be self-gravitating.

By ``statistical'' instability, we refer to the regime where an inhomogeneous disk experiences sufficiently large fluctuations in density such that there is an order-unity probability (integrated over the entire volume and lifetime of the disk) of the formation of {\em some} region which is so over-dense that it can successfully collapse under self-gravity. A ``statistically stable'' disk has a probability much less than unity of such an event occurring, even once in its lifetime. 

Classically unstable disks are always statistically unstable, and statistically stable disks are always classically stable. However, we argue in this paper that there is a large regime of parameter space in which disks can be {\em classically stable}, but {\em statistically unstable}. In this regime, disks can, in principle, evolve for millions of dynamical times with $Q\gg1$, and nearly all the disk mass will be stable, but rare density fluctuations might form an order-unity number of isolated self-gravitating ``fragments.''

\begin{figure*}
    \centering
    \plotsidesize{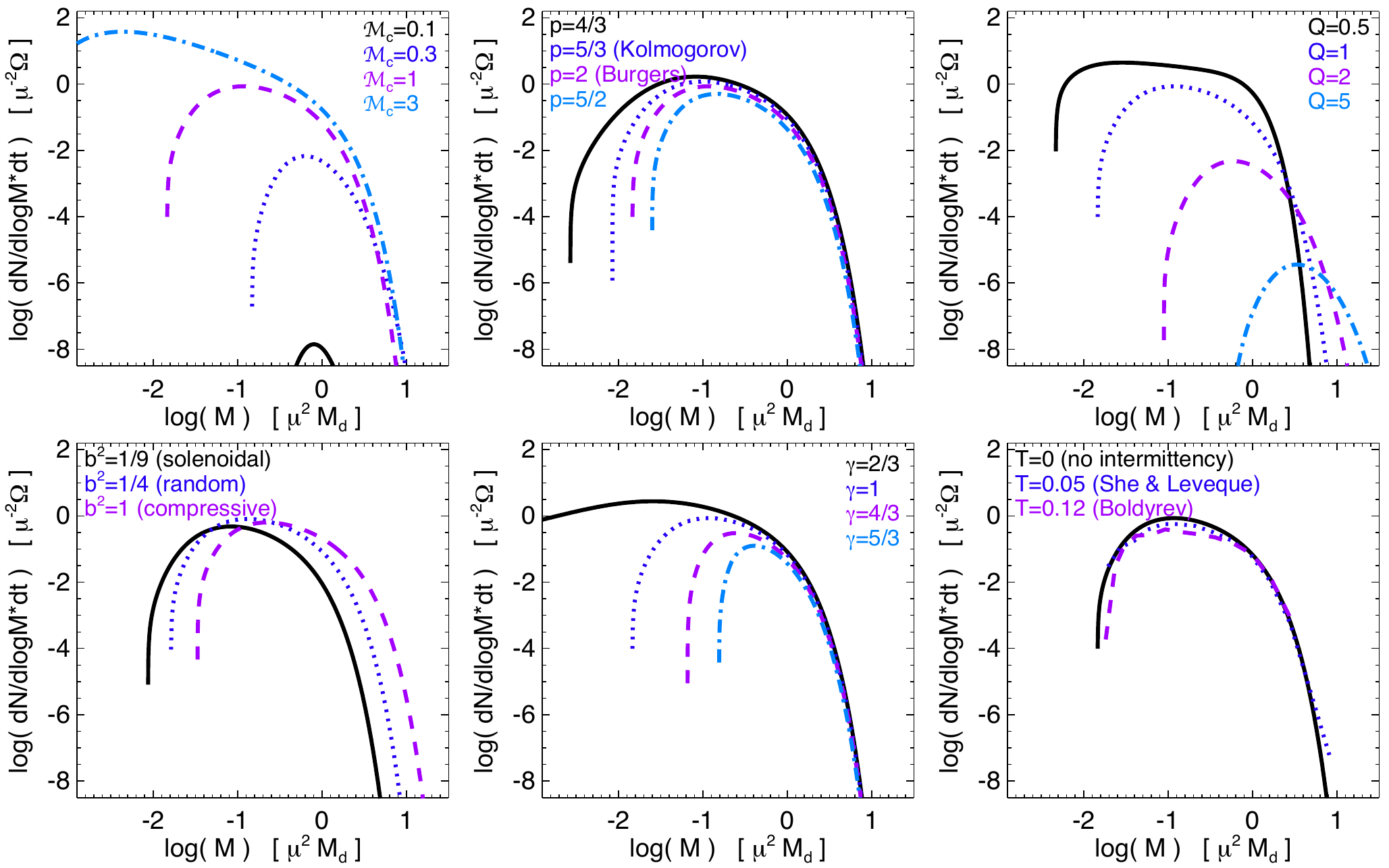}{0.95}
    \caption{Here we show the rate of formation of self-gravitating clumps as a function of mass. We plot the probability per unit time, per log-interval in mass, of the formation of a self-gravitating gas overdensity (turbulent density fluctuation) which can collapse -- i.e.\ a ``fragmentation event'' and candidate planet formation via direct collapse. Mass is in units of $\mu^{2}\,M_{d}$ where $M_{d}$ is the disk mass and $\mu\equiv M_{d}/M_{\ast}$ is the disk-to-star (central object) mass; the probability is in units of $\mu^{-2}\,\Omega$, where $\Omega = 2\pi/t_{\rm orbit}$ is the Keplerian frequency. We define a ``reference model'' ($\Machcompressive=1$, $p=2$, $Q=1$, $b=1/2$, $\gamma=1$, $T=0$) and vary each parameter in turn. 
    {\bf (1)} Large-scale compressive (longitudinal) Mach number $\Machcompressive = b\,\Mach[h]$. Fragmentation is exponentially suppressed when $\Machcompressive\ll1$, and develops on a single crossing time over a broad mass range when $\Machcompressive\gg1$. There is a characteristic mass $\sim\mu^{2}\,M_{\rm d}$.
    {\bf (2)} Turbulent spectral index ($E(k)\propto k^{-p}$), $p=2$ is Burgers (highly compressible) turbulence and $p=5/3$ is Kolmogorov (incompressible); values outside this range are rare. This has weak effects, but shallower spectra give more power on small scales.
    {\bf (3)} Global Toomre parameter ($Q\approx1$ for marginal classical stability). $Q\gg1$ exponentially suppresses (but does not eliminate) fragmentation. $Q\ll1$ leads to ``catastrophic'' fragmentation in a single crossing time on a broad range of mass scales.
    {\bf (4)} Fraction $b$ of turbulent velocity in compressive modes ($\Machcompressive=b\,\Mach$). In purely compressive turbulence $b=1$, pure solenoidal turbulence (and/or cases with strong magnetic fields) $b=1/3$, and random forcing $b=1/2$. At fixed compressive $\Machcompressive$ (not fixed $\Mach$), this has a weak effect.$^{\ref{foot:compressive.mach.correction}}$
    {\bf (5)} Equation of state polytropic index $\gamma$ ($c_{s}^{2}\propto\rho^{\gamma-1}$). ``Soft'' $\gamma<1$ produce a much broader spectrum of fragmentation on small scales, as compared to isothermal ($\gamma=1$). $\gamma=4/3$ corresponds to a radiation-pressure supported disk with no cooling; $\gamma=5/3$ to an adiabatic disk with no cooling. These cases suppress fragmentation on smaller scales, but still form collapsing regions via turbulent density fluctuations on larger scales with nearly the same integrated probability (normalization). 
    {\bf (6)} Intermittency parameter $T$ (see \papertwo; Appendix~B). $T=0$ is no intermittency; $T=0.05$ corresponds to the intermittency model of \citet{sheleveque:structure.functions}, appropriate for incompressible Kolmogorov turbulence; $T=0.12$ to the model of \citet{boldyrev:2002.structfn.model}, for more intermittent compressible super-sonic turbulence. 
    \label{fig:mfs}}
\end{figure*}

\begin{figure}
    \centering
    \plotonesize{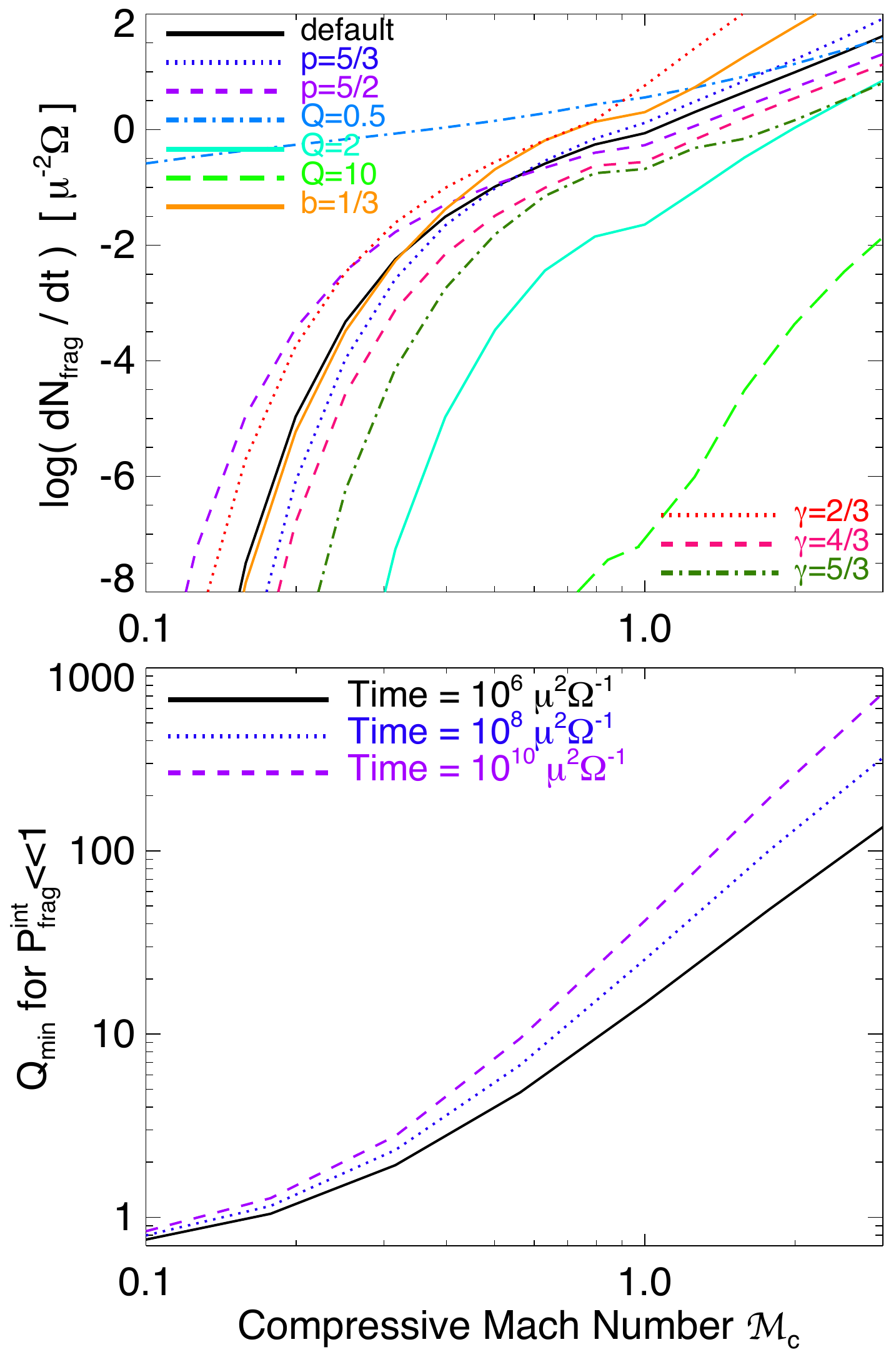}{0.99}
    \caption{{\em Top:} Probability (integrated over all masses in Fig.~\ref{fig:mfs}) per unit time of a fragmentation event vs.\ compressive Mach number $\Machcompressive$, for other varied parameters. 
    At fixed $\Machcompressive$ and $Q$, the turbulent power spectrum shape ($p$), forcing mechanisms and presence/absence of magnetic fields ($b$), equation-of-state ($\gamma$), and intermittency have small effects. In all cases the rate grows very rapidly when $\Machcompressive\gtrsim0.1$, and is suppressed when $Q\gg1$. 
    {\em Bottom:} The minimum value of $Q$ for statistical stability, as a function of $\Machcompressive$. For $Q>Q_{\rm min}$, the time-integrated probability of a fragmentation event $P_{\rm fragint}=\int {\rm d}t\,{\rm d}N_{\rm frag}/{\rm dt}$ is small (here we choose $\Pfragint<0.1$, but the exact choice has only a weak logarithmic effect on the result). The three lines span a plausible range for the ``lifetime'' of a turbulent protoplanetary disk (recall, $\Omega^{-1}\sim$\,yr at Jupiter, and typical $\mu\lesssim0.1$). The curves approximately follow $Q_{\rm min} \sim 
    0.5\,\exp{(\sqrt{2\ln{({\rm Time}/\mu^{2}\Omega^{-1})}\,\ln{(1+\Mach^{2})}})} \sim 
    0.5\,\exp{(6\,\sqrt{\ln{(1+\Mach^{2})}})}$ (see Eq.~\ref{eqn:Qmin}-\ref{eqn:betamin.maintext}).
    \label{fig:Qmin}}
\end{figure}

%\vspace{-0.5cm}
\section{Model Outline}

Here, we present an order-of-magnitude, qualitative version of the calculation which we will  develop rigorously below. This serves to illustrate some important scalings and physical processes. 

Consider an inhomogeneous disk (around a star of mass $M_{\ast}$) with scale radius $r_{\ast}$, scale-height $h$, mean surface density $\Sigma$ and mid-plane density $\rhodisk\approx \Sigma/(2\,h)$, and Toomre $Q > 1$. Although the disk is classically stable (and so not self-gravitating at the mean density), if a local ``patch'' of the disk exceeds some sufficiently large density then that region will collapse under self-gravity. The most unstable wavelengths to self-gravity are of order the disk scale height $\sim h$: roughly speaking, a gas parcel of this size will collapse under self-gravity if it exceeds the Roche criterion (overcomes tidal forces): $\rho \gtrsim M_{\ast}/r_{\ast}^{3} \sim Q\,\rhodisk$.\footnote{To derive this, we assume a Keplerian disk, $\Omega^{2}\approx G\,M_{\ast}/r_{\ast}^{3}$, and vertical equilibrium, $h\approx c_{s}/\Omega$ for a disk supported by thermal pressure.} 

So for $Q\gtrsim1$ we see that gas at the mean density will not collapse. However, turbulence produces a broad spectrum of density fluctuations. For simple, isothermal turbulence with characteristic (compressive) Mach number $\Mach = \langle v_{\rm turb}^{2} \rangle^{1/2} / c_{s}$, the (volumetric) distribution of densities is approximately log-normal, with dispersion $\sigma_{\ln{\rho}} \sim \sqrt{\ln[1 + \Mach^{2}]}$ ($\approx \Mach$ for $\Mach\lesssim1$). 

Since density fluctuations exceeding $\rho\gtrsim Q\,\rhodisk$ ($\ln{(\rho/\rhodisk)}\gtrsim \ln{Q}$) can collapse, we integrate the tail of the log-normal density probability distribution function (PDF) above this critical density to estimate the probability $P_{c}\sim {\rm erfc}[(\ln{Q})/(\sqrt{2}\,\sigma_{\ln{\rho}})]$, per unit volume, of a region being self-gravitating at a given instant. Since we are considering regions of size $\sim h$, the disk contains approximately $\sim (r_{\ast}/h)^{2}$ independent volumes, and so (assuming $P_{c}$ is small) the probability of any one such volume having a coherent volume-average density above the critical threshold is $P_{v}\sim (r_{\ast}/h)^{2}\,P_{c}$. 

Now consider that a typical proto-planetary disk (at a few AU) might have $Q\sim100$ and $\Mach\sim1$, with $h/r_{\ast} \sim 10^{-1}$. In this case, $P_{c}\sim 3\times10^{-8}$ and $P_{v}\sim 3\times10^{-6}$ are extremely small! 

However, this analysis applies to the disk viewed at a single instant. Turbulent density fluctuations evolve stochastically in time, with a coherence time (on a given scale) about equal to the bulk flow crossing time $\sim h/v_{t}$ (since this is the time for fluid flows to cross and interact with a new independent region of size $\sim h$). So the density PDF is re-sampled or ``refreshed'' on the timescale $\sim h/v_{t} \sim 1/(\Mach\,\Omega)$. But this is just a dynamical time, which is very short relative to a typical disk lifetime ($\Omega^{-1}\sim$\,yr at $\sim10$\,AU). If the disk survives for a total timescale $t_{0}$, then, the entire volume is re-sampled $\sim \Mach\,t_{0}\,\Omega$ independent times. So the probability, integrated over time, of any one of these volumes, at any one time, exceeding the self-gravity criterion, is $\Pfragint\sim \Mach\,t_{0}\,\Omega\,(r_{\ast}/h)^{2}\,{\rm erfc}[\ln{Q}/(\sqrt{2}\,\sigma_{\ln{\rho}})]$. 

If a typical disk with parameters above survives for $\sim$\,Myr, or $\sim 10^{6}$ crossing times, we then obtain a time-and-volume-integrated probability $\Pfragint \sim 1$ of at least a single stochastic ``fragmentation event'' driven by turbulent density fluctuations. The mass of the self-gravitating ``fragment'' will be $\sim (4\pi/3)\,h^{3}\,\rho \sim 4\,Q\,h^{3}\,\rhodisk$ {\bf $\sim 4\,(h/r_{\ast})^{3}\,M_{\ast}$} ($\sim 0.1-1\,M_{\rm Jupiter}$, for these parameters in a minimum mass solar nebula). And, despite the fact that the average timescale between such events may be long ($\sim$\,Myr), if a fragment forms, it forms rapidly (in $\sim$\,yr) on the turbulent crossing time $\sim 1/(\Mach\,\Omega)\sim$\,yr and has a short collapse/free-fall time $\sim 1/\sqrt{G\rho} \sim \Omega^{-1}\sim\,$yr. Despite having $Q\sim 100$, then, we estimate that this disk could be statistically unstable!

For otherwise fixed $h/r_{\ast}\sim0.1$ and $\Mach\sim1$, $\Pfragint$ declines exponentially with increasing $Q$. So for $Q\gg 100$, $\Pfragint\ll1$ and such a disk is both classically and statistically stable. For $Q<1$, the disk is classically unstable, and even material at the mean density (i.e.\ an order-unity fraction of the mass) begins collapsing on a single dynamical time. But for $1\lesssim Q \lesssim 100$, such a disk is classically stable, but statistically unstable. 

Below, we present a more formal and rigorous derivation of statistical (in)stability properties, and consider a range of disk properties. But the simple order-of-magnitude arguments above capture the most important qualitative behaviors we will discuss.

%\vspace{-0.5cm}
\section{The Model: Turbulent Density Fluctuations}
\label{sec:turb.density}

As mentioned above, turbulence in approximately isothermal gas (neglecting self-gravity) generically drives the density PDF to a approximate lognormal (a normal distribution in $\ln{(\rho)}$). This follows from the central limit theorem \citep[see][]{passot:1998.density.pdf,nordlund:1999.density.pdf.supersonic}, although there can be some corrections due to intermittency and mass conservation \citep{klessen:2000.pdf.supersonic.turb,hopkins:2012.intermittent.turb.density.pdfs}. And in the simplest case of an ideal box of driven turbulence, the variance $S$ simply scales with the driving-scale Mach number $\Mach$ as $S=\ln{[1+b^{2}\,\Mach^{2}]}$ (where $b$ is a constant discussed below). These scalings have been confirmed by a huge number of numerical experiments with $\Mach\sim0.01 - 100$, sampling the PDF down to values as low as $\sim10^{-10}$ in the ``tails'' \citep{federrath:2008.density.pdf.vs.forcingtype,federrath:2010.obs.vs.sim.turb.compare,federrath:2012.sfr.vs.model.turb.boxes,schmidt:2009.isothermal.turb,price:2010.grid.sph.compare.turbulence,konstantin:mach.compressive.relation}.

This is true in both sub-sonic and super-sonic turbulence (see references above), as well as highly magnetized media (with the magnetic fields just modifying $b$ or the effectively compressible component of $\Mach$; see \citealt{kowal:2007.log.density.turb.spectra,lemaster:2009.density.pdf.turb.review,kritsuk:2011.mhd.turb.comparison,molina:2012.mhd.mach.dispersion.relation}), and even multi-fluid media with de-coupled electrons, ions, neutral species, and dust \citep[see][]{downes:2012.multifluid.turb.density.pdf}.\footnote{{It is a common misconception that log-normal density PDFs apply only to super-sonic, non-magnetized turbulence. In fact, while they apply strictly to only isothermal (non-intermittent) turbulence (as discussed in \S~\ref{sec:turb.density}), the analytic derivation of lognormal PDFs actually assumes {small} (local) Mach numbers (see \citealt{nordlund:1999.density.pdf.supersonic}). The log-normal model (with the higher-order intermittency corrections we include in \S~\ref{sec:results}) and the simple assumptions we use for the scaling of the density power spectrum with velocity power spectrum have been tested in the sub-sonic limit in both simulations \citep[see][]{shaikh:2007.turb.density.weakly.compressible.evol,kowal:2007.log.density.turb.spectra,burkhart:2009.mhd.turb.density.stats,schmidt:2009.isothermal.turb,konstantin:mach.compressive.relation}, and also in experimental data from the solar wind \citep{burlaga:1992.multifractal.solar.wind.density.velocity,forman:2003.castaing.fits.solar.wind.data,leubner:2005.solar.wind.castaing.fits.density} and laboratory MHD plasmas \citep{budaev:2008.tokamak.plasma.turb.pdfs.intermittency} as well as jet experiments \citep{ruiz-chavarria:1996.passive.scalar.turb.poisson.tests,warhaft:2000.passive.scalar.turb.review,zhou:2010.scalar.field.logpoisson.pdfs}. The power spectrum predictions in this limit, in fact, generically follow the well-known and tested weakly-compressible Kolmogorov-like scaling \citep{montgomery:1987.density.fluct.powerspectra.trace.vel,zank:1990.nearly.incompressible.mhd.turbulence}. And in \citet{hopkins:2012.intermittent.turb.density.pdfs}, we show that deviations from lognormal statistics in the high-density wing of the distribution are less significant at lower Mach numbers. Likewise, because of the lower compressive Mach numbers, these assumptions are more accurate in solenoidally-forced and/or magnetized turbulence \citep[see][]{kowal:2007.log.density.turb.spectra,hopkins:2012.intermittent.turb.density.pdfs,federrath:2013.intermittency.vs.numerics}.}} And although the density PDF is not exactly lognormal when the gas is no longer isothermal, the inviscid Navier-Stokes equations show that the local response is invariant under the substitution $S(\Mach)\rightarrow S(\Mach\,|\,\rho) = \ln{[1+b\,v_{t}^{2}/c_{s}^{2}(\rho)]}$, allowing an appropriately modified PDF which provides an excellent approximation to the results in simulations \citep[see][]{scalo:1998.turb.density.pdf,passot:1998.density.pdf}. 

In \paperone-\papertwo, we use these basic results to show how excursion set theory can be used to {\em analytically} predict the statistical structure of turbulent density fluctuations. The details are given therein; for the sake of completeness, we include a summary of the important equations in Appendix~\ref{sec:appendix:model}. Here, we briefly describe the calculation. 

Consider the field $\rho_{{\bf x}}(R)=\rho({\bf x}\,|\,R)$: the density field about the (random) coordinate {\bf x} in space, smoothed with some window function of characteristic radius $R$ (e.g.\ the average density in a sphere). If the gas is isothermal (we discuss more complicated cases below), this is distributed as a log-normal: 
\be
P(\ln[\rho_{\bf x}(R)]) = \frac{1}{\sqrt{2\pi\,S(R)}}\,\exp{\left(-\frac{[\ln{\rho_{\bf x}(R)} + S(R)/2 ]^{2}}{2\,S(R)} \right)}
\ee
where $S(R)$ is the variance on each scale $R$. $S(R)$ is just the Fourier transform of the density power spectrum, which itself follows from the (well-defined) velocity power spectrum (see Eq.~\ref{eqn:S.R}). Essentially, $S(R)$ is determined by integrating the contribution to the variance on each scale from the velocity field, using the relation ${\Delta}S\approx \ln{(1 + b^{2}\,\mathcal{M}(R)^{2})}$ ($\mathcal{M}(R)$ is the scale-dependent Mach number of the velocity field). 

``Interesting'' regions are those where $\rho_{\bf x}(R)$ exceeds some critical value, above which the region is sufficiently dense so as to be self-gravitating. Including the effects of support from angular momentum/shear, thermal and magnetic pressure, and turbulence, this is given by 
\begin{align}
\label{eqn:rhocrit}
\frac{\rhocrit(R)}{\rhodisk} \equiv \frac{Q}{2\,\tilde{\kappa}}\,\left(1+\frac{h}{R} \right)
{\Bigl[} \frac{\sigma_{g}^{2}(R,\,\rhocrit)}{\sigma_{g}^{2}(h,\,\rhodisk)}\,\frac{h}{R}  + 
\tilde{\kappa}^{2}\,\frac{R}{h}{\Bigr]} 
\end{align}
\citep{vandervoort:1970.dispersion.relation}. Here $\rhodisk$ is the mean midplane density of the disk, $h$ is the disk scale height, $\tilde{\kappa}\equiv\kappa/\Omega=1$ for a Keplerian disk ($\kappa$ is the epicylic frequency), and 
$Q\equiv (\sigma_{g}[h,\,\rhodisk]\,\kappa)/(\pi\,G\,\Sigma_{\rm gas})$ is the Toomre $Q$ parameter. The effective gas dispersion $\sigma_{g}^{2}(R,\,\rho) = c_{s}^{2}(\rho) + \langle v_{t}^{2}(R) \rangle  + v_{\rm A}^{2}(\rho,\,R)$ (Eq.~\ref{eqn:sigmagas}) includes the thermal ($c_{s}$), turbulent ($v_{t}$), and magnetic support (Alfv{\'e}n speed $v_{\rm A}$). A full derivation is given in \paperone; but we stress that this not only implies/requires that a region with $\rho(R)>\rhocrit(R)$ is gravitationally self-bound, but {\em also} that such regions will not be unbound by tidal shear (i.e.\ have sizes within the Hill radius) and that they will not be unbound/destroyed by energy input from the turbulent cascade (shocks and viscous heating).\footnote{Note that in Eq.~\ref{eqn:rhocrit} we consider the rms turbulent velocity (i.e.\ do not explicitly treat different velocity fluctuations), even though we consider density fluctuations from the turbulence. In \paperone\ \&\ \papertwo\ we show that this is a very small source of error (see also e.g.\ \citealt{sheth:2002.linear.barrier}, who show that even for pressure-free flows with velocities typically above escape velocities, this introduces a $\sim10\%$ correction to the predicted mass function). But particularly here, since the turbulence of interest is sub-sonic, $\langle v_{t}^{2}(R) \rangle$ is {never} the dominant component of $\sigma_{g}^{2}$. Moreover, even though velocity fluctuations do drive the density fluctuations, because these are built up hierarchically in the cascade, the magnitude of the coherent velocity fluctuations on a given scale is instantaneously nearly uncorrelated with the density fluctations (see \citealt{federrath:2010.obs.vs.sim.turb.compare}, who find $\langle v_{t}(R) \rangle \propto \langle \rho(R)/\rho_{0} \rangle^{-0.05}$). Inserting such a scaling into our model is straightforward, but has no detectable effect in any figure shown. Finally, the most important effects of coupled velocity-density fluctuations are already included in the models for intermittency we consider, since this coupling makes the PDF non-lognormal.} Knowing the size $R$ and critical density $\rhocrit$ of a region, it is trivial to translate this to the total collapsing gas mass $M=M(R)$.

We desire the mass and initial size spectrum of regions which exceed $\rhocrit$ (so are self-gravitating) on the scale $R$ specifically defined as the largest scale on which the region is self-gravitating (i.e.\ excluding ``sub-units'' so that we do not double-count ``clouds within clouds''). In \paperone\ we show this reduces to a derivation of the ``first-crossing'' distribution for the field $\rho_{\bf x}(R)$. The mass function of collapsing objects can be written generally as 
\be
\frac{{\rm d}n}{{\rm d}M} = 
\frac{\rhocrit(M)}{M}\,\ffirst(M)\,{\Bigl |}\frac{{\rm d}S}{{\rm d}M} {\Bigr |}
\ee
where $\ffirst(S)$ is a function given in Eq.~\ref{eqn:ffirst} that is somewhat cumbersome to derive (see \paperone\ for details), but depends only on how the dimensionless quantities $S(R)$ and $\rhocrit(R)/\rhodisk$ ``run'' as a function of scale $R$.

In \papertwo, we further generalize this to fully time-dependent fields. In statistical equilibrium, the density field obeys a modified Fokker-Planck equation with different modes evolving stochastically -- they follow a damped random walk with a correlation time equal to the turbulent crossing time for each spatial scale/wavenumber. This allows us to directly calculate the probability per unit time of the formation of any bound object in the mass function above. The exact solution requires a numerical approach described in \paperone\ (\S~7) and \papertwo\ (\S~9-10), but to very good approximation (for $M\,{\rm d}n/{\rm d}\ln{M}\lesssim\rhodisk$) this is just 
\be
\Delta P(M,\,\Delta t)\approx \frac{{\rm d}n}{{\rm d}M}\,\frac{\Delta t}{\tau(M)}
\ee
where $\tau(M) = R[M] / v_{t}(R[M])$ is the turbulent crossing time on the scale $R[M]$ corresponding to the mass (Eq.~\ref{eqn:mass.radius}). This is because the coherence time of density fluctuations on a given scale is just the crossing time. 

Now consider a disk, or disk element (if we consider a series of cylindrical annuli at different disk-centric radii) with total mass $M_{d}$. By the definitions used in \paperone-\papertwo, this means the total ``effective volume'' is $M_{d}/\rhodisk$, i.e.\ that the total (integrated) number of objects per unit mass is 
\be
\frac{{\rm d}N}{{\rm d}\log{M}} = \frac{M_{d}}{\rhodisk}\,\frac{{\rm d}n}{{\rm d}\log{M}} 
= \frac{M_{d}}{M}\,\frac{\rhocrit}{\rhodisk}\,\ffirst(M)\,{\Bigl|}\frac{{\rm d}S}{{\rm d}\log{M}}{\Bigr|}
\ee

For any polytropic gas, we can factor out all dimensional parameters, and work in units of $h$ and $\rhodisk$. Mass then has units of $\rhodisk\,h^{3}$. But since, for a disk in vertical equilibrium $h=\sigma_{g}/\Omega$, and for a Keplerian disk $\kappa = \Omega = (G\,M_{\ast}\,r_{\ast}^{-3})^{1/2}$ (where $M_{\ast}$ is the central mass and $r_{\ast}$ is the disk-centric radius), we have $Q =(\sigma_{g}[h]\,\kappa)/(\pi\,G\,\Sigma_{\rm gas}) = h\,\Omega^{2}/(\pi\,G\,\Sigma_{\rm gas}) = (h/r_{\ast})\,M_{\ast}/(\pi\,\Sigma_{\rm gas}\,r_{\ast}^{2}$). So we can write $(h/r_{\ast}) = Q\,\mu$ where 
\be
\mu \equiv \frac{\pi\,\Sigma_{\rm gas}\,r_{\ast}^{2}}{M_{\ast}} \approx \frac{M_{d}}{M_{\ast}}
\ee
where $M_{d}$ is the disk mass within $R$. For an exponential vertical profile used to define our dispersion relation, $\Sigma_{\rm gas}=2\,\rhodisk\,h$, so we can use this and $(h/r_{\ast})=Q\,\mu$ to link $\rhodisk\,h^{3} = (2\pi)^{-1}\,(\mu\,Q)^{2}\,M_{d}$. We can then remove the local quantities $\rhodisk$ and $h$ and define mass in the much simpler global units of $\mu$ and $M_{d}$. The units of ${\rm d}N$ defined above are $M_{d}/(\rhodisk\,h^{3})$, so this can be similarly re-written. And since the disk crossing time scales as $\sim h/\sigma_{g}[h] = \Omega^{-1}$, $\Omega$ provides a natural time unit. 

Having defined units, the model is completely specified by dimensionless parameters. These are the spectral index $p$ of the turbulent velocity spectrum, $E(k)\propto k^{-p}$ ($v_{t}^{2}(R)\propto R^{p-1}$), and its normalization, which we define by the Mach number on large scales $\Mach_{h}^{2}\equiv \langle v_{t}^{2}(h)\rangle/c_{s}^{2}$, as well as the Toomre parameter $Q$. We must also specify the parameter $b$, i.e.\ the mean fraction of the velocity in compressive modes. This is $b=1$ for purely compressively forced turbulence, $b=1/3$ for purely solenoidally forced turbulence, and $b=1/2$ for random forcing. But we can almost completely factor out the dependence on this parameter if we simply define the compressive component of the turbulence, i.e.\ work in units of the {\em compressive} Mach number\footnote{\label{foot:compressive.mach.correction}{At sufficiently small $\Machcompressive$, the compressive-to-total Mach number ratio can scale steeply with Mach number depending on the turbulent forcing \citep[see][]{zank:1990.nearly.incompressible.mhd.turbulence,zank:1990.solar.wind.density.fluctuations}. Because we work specifically with the compressive Mach number, this is largely irrelevant to our calculation (what {\em does} matter is that the relation between $\Machcompressive$ and density fluctuations remains intact; see \citealt{konstantin:mach.compressive.relation}). Moreover the ``steepening'' becomes significant only below the minimum $\Machcompressive$ values we will identify as interesting (compare e.g.\ Fig.~6 of \citealt{konstantin:mach.compressive.relation}), especially for magnetized turbulence (where other effects have the opposite sense; see \citealt{ostriker:2001.gmc.column.dist,shaikh:2007.anti.correlated.thermal.density.fluctuations,price:2011.density.mach.vs.forcing,molina:2012.mhd.mach.dispersion.relation}).}} $\Machcompressive\equiv b\,\Mach$.

In \papertwo, we show how these equations generalize for the cases with non-isothermal gas, intermittent turbulence, and non-isotropic magnetic fields. The qualitative scalings are similar, but the math becomes considerably more complicated (and the ``first-crossing distribution'' $\ffirst(M)$ and its evolution in time must be solved via a numerical Monte Carlo method rather than analytically). The presence of intermittency makes the density PDF non-Gaussian, and introduces explicitly correlated fluctuation structures, but this can be entirely encapsulated in a modified form of Eq.~\ref{eqn:ffirst} above (see \papertwo, Appendix~D), leaving the rest of our derivation intact. We will consider such non-Gaussian, correlated statistics and show they give very similar results.
For non-isothermal cases, we must replace $c_{s}\rightarrow c_{s}(\rho)$ (in calculating both the variance and critical density), and again allow for the density PDF to be non-Gaussian (with a skew towards lower/higher densities as the equation of state is made more or less ``stiff,'' respectively; see \S~3-4 in \papertwo; this also introduces higher-order correlations in the fluctuation statistics, distinct from those associated with intermittency). Modulo these changes, however, our derivations (and the changes made to specify to the Keplerian disk case) are identical for the simple case of a polytropic gas where $c_{s}^{2} \propto \rho^{\gamma-1}$ ($\gamma$ is the polytropic index). Magnetic fields can produce global anisotropy, but this can be simply absorbed into the form of $\ffirst(M)$ as well; their largest effect in suppressing fluctuations manifests as a field-strength-dependent value $b$ (\papertwo, \S~6, and e.g.\ \citealt{kowal:2007.log.density.turb.spectra,lemaster:2009.density.pdf.turb.review,molina:2012.mhd.mach.dispersion.relation}). For the strong-field limit, however, this is just similar to the pure-solenoidal turbulence case (in both cases, there is only one spatial dimension along which compression is possible); so is within the range of $b$ variations we will consider.\footnote{We do caution that some of our simple assumptions (for example, how we assume the density and velocity power spectra ``turn over'' at large scale heights $z\gtrsim h$) remain to be tested in simulations of fully nonlinear, shearing, vertically stratified, MHD turbulence. However, within the disk scale height, preliminary comparisons suggest these corrections have little effect. In \S~\ref{sec:mri} below, we explicitly compare the predictions from our model to the variance in the mid-plane density field calculated in such simulations (vertically stratified, shearing MRI boxes; \citealt{bai:2012.mri.saturation.turbulence}), and find our predictions agree remarkably well over a range of field strengths. Moreover, we can repeat this comparison at various vertical heights up to $\sim 6\,h$, and find $\sim10\%$ agreement. Recently, \citet{arena:2013.protoplanetary.disk.fluctuations.sph} have performed numerical experiments of three-dimensional (non-magnetized, isothermal) disks with effective Mach numbers from $\sim1$ to $<0.1$ ($\Machcompressive \lesssim0.03$); the results agree well with our assumed form for the density distribution and power spectrum shape. In future work (J.\ Lynn et al., in preparation), we will study the effect of non-isothermal, stratified, rotating, and self-gravitating flows on turbulent density fluctuations.}
\section{Results: Fragmentation Rates and Mass Spectra in the General Case}
\label{sec:results}

In Fig.~\ref{fig:mfs}, we now use this to predict the mass spectrum -- specifically the probability per unit time, per unit mass -- of the formation of self-gravitating regions, as a function of various properties of the system.\footnote{For simplicity, we will refer to these calculations as if the disk is constant surface-density out to some maximum radius, with total mass $M_{d}$ and $\Omega$ (the orbital velocity) defined at that radius, and constant Toomre $Q$ parameter. However, the results shown could be applied to any radius in a disk with any mass profile (provided it is still Keplerian), with $M_{d}$ defined as the disk mass inside a radial annulus, and $Q$ and $\Omega$ evaluated at that same radius.} 

We define our ``reference'' model to be isothermal ($\gamma=1$), non-intermittent (i.e.\ log-normal density PDF), with turbulent spectral slope $p=2$ (appropriate for compressible turbulence), Toomre $Q=1$, and have $b=1/2$ (random forcing) with rms {\em three-dimensional} compressive Mach number on the largest scales $\Machcompressive=1$. But we then vary all of these parameters. For now, we treat each as free -- in other words, we make no assumptions about the specific mass profile, temperature structure, cooling chemistry, or other microphysics of the disk. These microphysics are, of course, what ultimately determine the value of the model parameters (and may build in some intrinsic correlations between them); but for any given set of parameters in Fig.~\ref{fig:mfs}, the prediction is independent of {\em how} the microphysics produce those parameters. 

In general, the {\em shape} of the mass spectrum is similar regardless of these variations. This is  peaked (very approximately lognormal-like) around a characteristic mass $\sim0.1-1\,\mu^{2}\,M_{d}$. For a relatively massive disk $M_{d}\sim0.1\,M_{\ast}$, then, we expect characteristic masses in such events of $\sim10^{-3}\,M_{\ast}$, corresponding to gas giants. If however the disk mass is lower, $M_{d}\sim0.01\,M_{\ast}$, then this becomes $\sim 10^{-6}\,M_{\ast}$, typical of rocky (Earth-mass) planets.

The dependence on the slope of the turbulent spectrum is quite weak: Kolmogorov-like ($p=5/3$, appropriate for incompressible turbulence) spectra give nearly identical results; shallower spectra slightly broaden the mass range predicted (since $\Machcompressive$ declines more slowly at small scales), but these are not seen in realistic astrophysical contexts. Likewise, at fixed $\Machcompressive$, there is some dependence on $b$ (because of how $\Mach = \Machcompressive/b$ enters into the critical density for collapse), but this is small in this regime, because turbulence is not the dominant source of ``support'' resisting collapse. The effects of intermittency are very weak, since they only subtly modify the density PDF shape (see \papertwo; {the important point here is that our predictions do not much change for reasonable departures from Gaussian or log-normal statistics}). Even changing the equation of state has surprisingly mild effects. A ``stiffer'' (higher-$\gamma$) equation of state is more resistive to fragmentation on small scales and leads to a more sharply peaked spectrum. But for fixed $\Machcompressive$, the variance near the ``core'' of the density PDF is similar independent of $\gamma$, so this does not have much effect on the normalization of the probability distribution. We stress that the medium could have a perfectly adiabatic equation of state with no cooling, and {\em if} it had the plotted Mach number and Toomre $Q$, our result would be identical (and the probability of fragmentation would still be finite). Very soft equations of state, on the other hand, can lead to a runaway tail of small-scale fragmentation, but this is not likely to be relevant. Clearly, the largest effects come from varying $\Machcompressive$ and $Q$; at $\Machcompressive\gtrsim1$ or $Q\lesssim1$ the mass distribution rapidly becomes more broad, while at $\Machcompressive\lesssim1$ or $Q\gtrsim1$ the characteristic mass remains fixed but the normalization of the probability becomes exponentially suppressed. 

This is summarized in Fig.~\ref{fig:Qmin}. Since the mass range of expected ``events'' is relatively narrow, we integrate over mass to obtain the total probability of an event per unit time, 
\be
\frac{{\rm d}N_{\rm frag}}{{\rm d}t} = \int\,{\rm d}\log{M}\,\frac{{\rm d}N}{{\rm d}\log{M}\,{\rm d}t}
\ee
and plot this as a function of $\Machcompressive$ for different parameter choices. As before, most parameters make a surprisingly small difference; $\Machcompressive$ and $Q$ dominate.

%\vspace{-0.5cm}
\subsection{A General Statistical Stability Criterion}
\label{sec:results:stability}

Formally ${\rm d}N_{\rm frag}/{\rm d}t$ is always non-zero for $\Machcompressive>0$; but we see that for for $\Machcompressive \lesssim 1/2$, the probability per unit time of forming a self-gravitating fluctuation drops rapidly.
However, recall that the total lifetime of e.g.\ a proto-planetary disk is many, many disk dynamical times $\sim \Omega^{-1}$, and our ``time unit'' is $\mu^{2}\,\Omega^{-1}$. Consider a typical lifetime of $\tau_{\rm Myr} \equiv \tau_{\rm disk}/{\rm Myr} \sim 1$; then the disk at $\sim 10$\,AU around a solar-mass star (where $\Omega\approx 1\,{\rm yr^{-1}}$) experiences $\sim 10^{6}$ dynamical times. For a disk-to-total mass ratio of $\mu\sim0.1$, this is $\sim10^{8}$ ``time units,'' so if we integrate the probability of a fragmentation event over the lifetime of the disk we obtain an order-unity probability even for ${\rm d}N_{\rm frag}/{\rm d}t\sim 10^{-8}$ in the units here (i.e.\ $\Machcompressive$ as small as $\sim0.15$). Of course, following such an integration in detail requires knowing the evolution of the disk mass, $Q$, $\Machcompressive$, etc. But we can obtain some estimate of the value of $Q$ required for statistical stability (ensuring the probability of fragmentation events is negligible) by simply assuming all quantities are constant and integrating over an approximate timescale, also shown in Fig.~\ref{fig:Qmin}. Here we  take our standard model, and consider three timescales in units of $\mu^{2}\,\Omega^{-1}$ (a factor of $\sim100$ shorter and longer than the value motivated above), and consider the minimum $Q$ at each $\Machcompressive$ needed to ensure that the time-integrated probability of a fragmentation event is $\ll1$.

This minimum $Q_{\rm min}$ is $\sim1$ at $\Mach\sim0.1$; equivalently, disks with $Q\approx1$ and $\Machcompressive\gtrsim0.1$ have an order-unity probability of at least one stochastic fragmentation event over their lifetime ($\Pfragint\sim1$). At larger $\Machcompressive\gtrsim0.3-0.5$, $Q\gtrsim 3-5$ is required for $\Pfragint\ll1$; and by sonic Mach numbers $\Machcompressive\sim1-3$, $Q\gtrsim40-1000$ is required for $\Pfragint\ll1$. 

We can approximate the scaling of $Q_{\rm min}(\Machcompressive)$ by the following: recall that the critical density near the Toomre scale $h$ scales approximately as $\ln{(\rhocrit/\rho_{0})}\sim \ln{(2\,Q)}$ (Eq.~\ref{eqn:rhocrit}), while the density dispersion scales as $\sigma_{\ln{\rho}}\sim\sqrt{(1+\Machcompressive^{2})}$ (Eq.~\ref{eqn:S.R}). As noted above, if the system evolves for a total timescale $\tau_{0} = t_{0}/(\mu^{2}\,\Omega^{-1})$ (time in our dimensionless units), then an event with probability per unit time $P \approx 1/\tau_{0}$ has an order-unity probability of occurring. If the probabilities are approximately normally distributed then this is just $\exp(-B^{2}/(2\,S)) \approx 1/\tau_{0}$, where $B$ is the barrier and $S$ the variance. Since the mass function is peaked near the Toomre scale we can approximate both by their values near the ``driving scale'' $\sim h$, $B\approx\ln{(2\,Q)}$ and $S\approx\ln{(1+\Machcompressive^{2}})$. Thus, statistical stability over some timescale of interest $t_{0}$ requires a $Q_{\rm min}$ in Fig.~\ref{fig:Qmin} of
\begin{align}
\label{eqn:Qmin} 
Q_{\rm min} &\approx 0.5\,\exp{[\sqrt{2\,\ln{(t_{0}\,\mu^{-2}\,\Omega)}\,\ln{(1+\Machcompressive^{2})}}]}  
\end{align}

For typical values of $t_{0}$, $\Omega$, and $\mu$ -- i.e.\ the typical number of independent realizations (in both time and space) of the turbulent field in a protoplanetary disk, this becomes
\begin{align}
Q_{\rm min}\sim 0.5\,\exp{(6\,\sqrt{\ln{(1+\Machcompressive^{2})}})}
\end{align}
In other words, a $\sim5-6\,\sigma_{\ln{\rho}}$ event has order-unity chance of occuring once over the disk lifetime, so for any $\Machcompressive$ this implies a minimum $Q$ needed to ensure statistical stability in such an extreme event.

\begin{figure}
    \centering
    \plotonesize{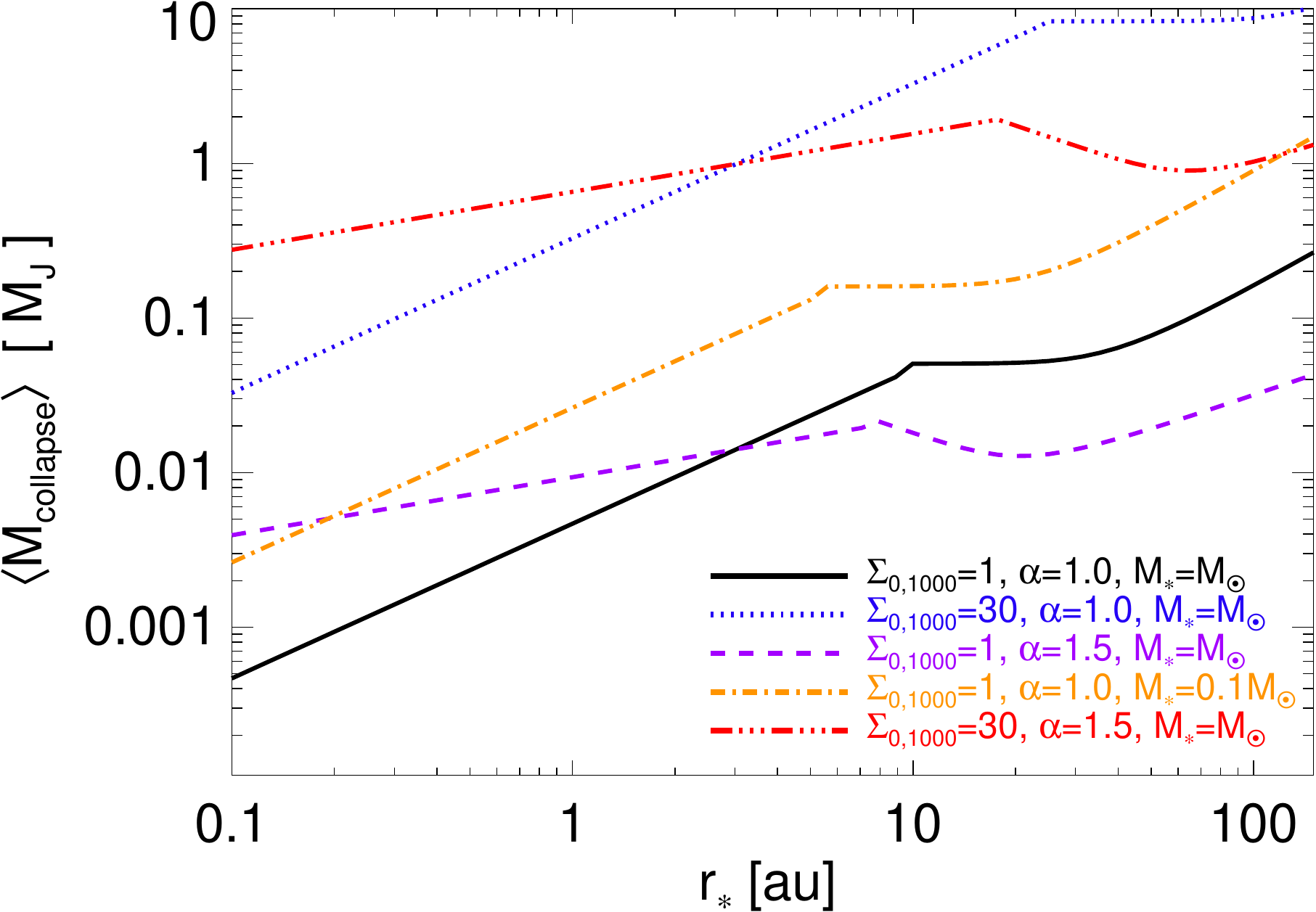}{1.01}
    \caption{Characteristic mass at collapse -- i.e.\ ``seed'' or ``collapse'' mass -- as a function of radius, for a protoplanetary disk with surface density profile $\Sigma = \Sigma_{0,\,1000}\,{1000\,{\rm g\,cm^{-2}}}\,(r_{\ast}/{\rm au})^{-\alpha}$, around a star with mass $M_{\ast}$ and disk temperature calculated including illumination and accretion as described in \S~\ref{sec:example.model}. Specifically, the mass is the mean $\langle M \rangle$ of the predicted MF as in Fig.~\ref{fig:mfs}, calculated with $Q$ for this temperature and $\Sigma$, and turbulent $b=1/2,\,T=0,\,\gamma=7/5,\,p=2,\,\Machcompressive=0.1$ (these parameters have little effect on the prediction). The MMSN corresponds to $\Sigma_{0,\,1000}=1$, $\alpha=1.5$, $M_{\ast}=M_{\sun}$. Units are Jupiter masses. The characteristic mass scales as $\sim \mu^{2}\,Q^{2}\, M_{d}(<r_{\ast})$. This (on average) increases with $r_{\ast}$, spanning an Earth-to-Jupiter mass range ($M_{\rm Earth}=0.003\,M_{\rm J}$).
    \label{fig:mass.vs.r}}
\end{figure}

%\vspace{-0.5cm}
\section{How Is the Turbulence Powered? Statistical Stability in Specific Models for Turbulence}
\label{sec:specific}

Thus far, we have considered the general case, varying the Mach numbers $\Machcompressive$ independent of other disk properties such as $Q$, $\Sigma$, and $\gamma$. However, in a realistic physical model, the mechanisms that drive turbulence may be specifically tied to these properties. Moreover, there may be certain characteristic Mach numbers expected or ruled out. In this section, we therefore consider some well-studied physical scenarios for the driving of turbulence in Keplerian disks, and examine their implications for the ``statistical stability'' we have described above.

%\vspace{-0.5cm}
\subsection{The ``Gravito-turbulent'' Regime (Gravity-Driven Turbulence)}
\label{sec:beta.disks}

Much of the work studying fragmentation in Keplerian disks has considered disks with a (locally) constant-cooling rate (``$\cbeta$'' disks, with $t_{\rm cool} = \cbeta\,\Omega^{-1}$ locally fixed). In particular, this includes the scenario of a ``gravito-turbulent'' steady-state from \citet{gammie:2001.cooling.in.keplerian.disks}, with local instabilities (spiral waves) powering turbulence which contributes an effective viscosity and maintains a steady temperature and $Q\sim1$. The theory we present above is more general than this: we make no assumption about the detailed cooling physics, or that the disk is an $\alpha$-disk, and allow the various parameters $Q$, $\Machcompressive$, $\gamma$, etc.\ to freely vary, whereas many of these are explicitly linked in these models. But in the theory above we cannot predict these quantities (or their co-dependencies); therefore, this model provides a simple and useful way to relate and predict some of the otherwise independently free parameters of the more general case, and is worth considering in detail.

%\vspace{-0.5cm}
\subsubsection{General Scalings}
\label{sec:beta.disks:general}

Consider a cooling rate which is uniform over a region (annulus) of the disk, 
\be
t_{\rm cool}=\cbeta\,\Omega^{-1}
\ee
If dissipation of gravitational instabilities (e.g.\ spiral waves) provides a source of heating balancing cooling, and $\cbeta\gtrsim3$, the system can develop a quasi steady-state angular momentum transport and Toomre $Q$ parameter, as in a \citet{shakurasunyaev73} $\alpha$-disk. \citet{pringle:accretion.review} and \citet{gammie:2001.cooling.in.keplerian.disks} showed that in this equilibrium, the ``effective'' viscosity parameter $\alpha$
\be
\alpha\approx\frac{4}{9}\,\frac{1}{\gamma\,(\gamma-1)\,\cbeta}
\ee
is approximately constant. This corresponds to the amplitude of density waves $\delta\Sigma/\Sigma \propto \alpha^{1/2}$, leading to a maximum $\alpha_{\rm max}\approx0.06$ above (hence minimum $\cbeta_{\rm min}$ below) which the local $Q<1$ and catastrophic fragmentation will occur \citep[see][]{gammie:2001.cooling.in.keplerian.disks,rice:2005.disk.frag.firstlook,cossins:2009.grav.instab.vs.coolingrate}. In an $\alpha$-disk, the inflow rate is also determined, as 
\be
\dot{M}=3\pi\,\alpha\,c_{s}^{2}\,\Sigma_{\rm gas}\,\Omega^{-1}
\ee
so this also corresponds to a maximum ``classically stable'' inflow rate below which catastrophic fragmentation will not occur. 

Implicitly, the relations above also define a steady-state Mach number. Recall, the dissipation of spiral instabilities is ultimately governed by the turbulent cascade. Since the turbulent dissipation rate is constant over scale in a Kolmogorov cascade, we can take it at the top level, ${\rm d}E/{\rm d}A\,{\rm d}t = (p-1)^{-1}\,\Sigma_{\rm gas}\,v_{t}^{2}\,\Omega$ (where here we consider the rate per unit area) and equate it to the cooling rate $=[\gamma\,(\gamma-1)]^{-1}\,\Sigma_{\rm gas}\,c_{s}^{2}\,t_{\rm cool}^{-1}$, giving $\Machcompressive^{2} \approx (3/2)\,\alpha$. Equivalently we could have equated the Reynolds stress that leads to $\alpha$, $\alpha=({\rm d}\ln{\Omega}/{\rm d}\ln{r_{\ast}})^{-1}\,T_{{\rm Rey}}/(\Sigma_{\rm gas}\,c_{s}^{2})$ with $T_{{\rm Rey}} = \langle \Sigma_{\rm gas}\,\delta v_{r} \delta v_{\phi}\rangle$; we obtain
\be
\label{eqn:mach.alpha}
\Machcompressive^{2} \approx \frac{3}{2}\,\alpha \approx \frac{2}{3}\,\frac{1}{\gamma\,(\gamma-1)\,\cbeta}
\ee
again.\footnote{One subtlety here is that the hydrodynamic Reynolds stress, and/or the dissipation on small scales, is dominated by the longitudinal (compressive) component. So this relation actually determines $\Machcompressive$, not necessarily $\Mach$. But for our purposes, this is particularly convenient, as it allows us to drop the $b$ term from our earlier derivation.} We now see how this relates to the theory developed in this paper. 

In the language here, increasing $\cbeta$ enters our theory by -- as discussed in the text -- changing the equilibrium balance between thermal and turbulent energy, i.e.\ the Mach number. As cooling becomes less efficient, maintaining the same $Q\sim1$ needed to power the turbulence (since spiral waves are still generated) requires a smaller turbulent dispersion, hence makes the system ``more stable.'' 

%\vspace{-0.5cm}
\subsubsection{A Statistical Stability Criterion}
\label{sec:beta.disks:stability}

But this also suggests an improved statistical stability criterion, accounting not just for regions where $Q<1$ but {\em also} for stochastic local turbulent density fluctuations. For a given $Q$, we have calculated (Fig.~\ref{fig:Qmin}) the Mach number $\Machcompressive$ above which the system will be probabilistically likely to fragment on a given timescale. Eq.~\ref{eqn:mach.alpha} allows us to translate this to a minimum $\cbeta$. We can do this exactly by simply reading off the numerically calculated values, but we can also obtain an accurate analytic approximation by the following (see \S~\ref{sec:discussion}). Recall that for a system which evolves for a total timescale $\tau_{0} = t_{0}/(\mu^{2}\,\Omega^{-1})$ (time in the dimensionless units we adopt), we obtain the approximate Eq.~\ref{eqn:Qmin} for the $Q_{\rm min}$ needed to ensure $\Pfragint\ll1$:
\be
Q \ge\frac{1}{2}\,\exp{[\sqrt{2\,\ln{(\tau_{0})}\,\ln{(1+\Machcompressive^{2})}}]} 
\ee
We can invert this to find the maximum $\Machcompressive$ for statistical stability at a given $Q$:
\be
\Machcompressive^{2} \lesssim \exp{{\Bigl[} \frac{[\ln{(2\,Q)}]^{2}}{2\,\ln{\tau_{0}}} {\Bigr]}} - 1 
\approx \frac{[\ln{(2\,Q)}]^{2}}{2\,\ln{\tau_{0}}}
\ee
Where the second equality follows from the fact that (for the systems of interest) $2\,\ln{(\tau_{0})}\gg\ln{(2\,Q)}$ is almost always true. 

Combining this with Eq.~\ref{eqn:mach.alpha}, we obtain:
\begin{align}
\label{eqn:Qmin.gravitoturb}
\cbeta_{\rm min} &\approx \frac{4}{3\,\gamma\,(\gamma-1)}\,\frac{\ln{(\tau_{0})}}{[\ln{(2\,Q)}]^{2}} \\ 
&= \frac{4}{3\,\gamma\,(\gamma-1)}\,\frac{\ln{[t_{0}/(\mu^{2}\,\Omega^{-1})]}}{[\ln{(2\,Q)}]^{2}}
\end{align}

We can immediately see some important consequences. Because of the stochastic nature of turbulent density fluctuations, $\cbeta_{\rm min}$ will {\em never} converge in time integration (assuming the disk can maintain steady-state mean parameters) -- there is always a finite (although possibly extremely small) probability of a strong shock or convergent flow forming a region which will collapse rapidly. However, the divergence in time is slow (logarithmic). The critical $\cbeta$ also scales with $\gamma$ just as in the \citet{gammie:2001.cooling.in.keplerian.disks} case; as the equation of state is made ``stiffer,'' the Mach numbers and density fluctuations are suppressed so faster cooling (lower $\cbeta$) can be allowed without fragmentation. And $\cbeta$ scales inversely with $\log{(Q)}$, so indeed higher-$Q$ disks are ``more stable,'' but there is no ``hard'' cutoff at a specific $Q$ value.

We can turn this around, and estimate the typical timescale for the formation of an order-unity number of fragments at a given $\cbeta$, obtaining
\be
\langle t(N_{\rm frag}\sim1) \rangle \approx \mu^{2}\,\Omega^{-1}\,\exp{{\Bigl(}\frac{3}{4}\,\cbeta\,\gamma\,(\gamma-1)\,[\ln{(2\,Q)}]^{2} {\Bigr)}}
\ee
As expected, this quickly becomes large for modest $\cbeta$ and/or $Q$. We stress, however, that this is a {\em probabilistic} statement. Although the mean timescale between fragment formation events might be millions of dynamical times, if and when individual fragments form meeting the criteria in the text, they do so rapidly -- on of order a single crossing time.

Given our derivation of $\cbeta_{\rm min}$, what do we expect in realistic systems such as proto-planetary disks? For a physical disk with $\cbeta\gg1$ we should expect $\gamma\approx7/5-5/3$ (depending on the gas phase), and in equilibrium $Q\sim1$; and we should integrate over the entire lifetime of the disk, $\tau_{0} = \tau_{\rm tot} \sim 10^{6}-10^{10}$ (with values motivated in \S~\ref{sec:results}). We then expect 
\be
\cbeta_{\rm min}^{\rm int} \approx \frac{51}{\gamma\,(\gamma-1)}\,\frac{1+0.05\,\ln{(\tau_{0}^{\rm int}/10^{8})}}{(1+1.4\,\ln{Q})^{2}}
\ee
This is fairly sensitive to $Q$ -- note $\cbeta_{\rm min}^{\rm int}\approx 13/[\gamma(\gamma-1)]$ if $Q=2$ instead -- and weakly sensitive to $\tau_{0}$ for reasonable variations. But this implies that only disks with {\em extremely} slow cooling, $t_{\rm cool}\gtrsim 50\,\Omega^{-1}$, (corresponding to steady-state Mach numbers $\Machcompressive\lesssim0.02$) are truly statistically stable with $Q\approx1$ over such a long lifetime.

%\vspace{-0.5cm}
\subsubsection{Comparison with Simulations}
\label{sec:beta.disks:sims}

Now consider the parameter choices in some examples that have been simulated. \citet{gammie:2001.cooling.in.keplerian.disks} considered the case with $\gamma=2$, and a steady-state $Q\approx 2.46$, evolving their simulations for typical timescales $t_{0}\sim50\,\Omega^{-1}$ (though they consider some longer-scale runs). Because these were two-dimensional shearing-sheet simulations, the appropriate $\mu$ is somewhat ambiguous, but recall $(h/r_{\ast}) = Q\,\mu$ by our definitions, and for the assumptions in \citet{gammie:2001.cooling.in.keplerian.disks} their ``standard'' simulation corresponds to an $h/r_{\ast}\approx 0.01\,Q$ (where we equate the ``full disk size'' to the area of the box simulated). Plugging in these values, then, we predict $\cbeta_{\rm min} = 3.4$, in excellent agreement with the value $\cbeta\approx3$ found by trial of several values therein. \citet{paardekooper:2012.stochastic.disk.frag} considered very similar simulations but with $Q\approx1$ and all sheets run for $t_{0}\sim1000\,\Omega^{-1}$; for this system we predict $\cbeta_{\rm min} = 20.4$ -- again almost exactly their estimated ``fragmentation boundary.'' \citet{meru:2012.nonconvergence.disk.frag} and \citet{rice:2012.convergence.disk.frag} consider three-dimensional global simulations; here $\mu=M_{d}/M_{\ast}=0.1$ is well-defined, a more realistic $\gamma=5/3$ is adopted, and the disks self-regulate at $Q\approx1$; the simulations are run for a shorter time $\sim50-100\,\langle\Omega\rangle^{-1}$ (where for convenience we defined $\langle\Omega\rangle$ at the effective radius of the disk, since it is radius-dependent, but this is where the mass is concentrated), giving $\cbeta_{\rm min}\approx21-25$, in very good agreement with where both simulations appear to converge (using either SPH or grid-based methods). This is also in good agreement with the earlier simulations in \citet{rice:2005.disk.frag.firstlook}, for $\gamma=5/3$ (predicting $\cbeta_{\rm min}=7.5$, vs.\ their estimated $6-7$) and $\gamma=7/5$ (predicting $\cbeta_{\rm min}=14$, vs.\ their estimated $13$). Of course, we should naturally expect some variation with respect to the predictions, since this is a stochastic process, but we do not find any highly discrepant results.

We should also note that convergence in the total fragmentation rate in simulations -- over any timescale -- requires resolving the full fragmentation mass distribution in Fig.~\ref{fig:mfs}. Unlike time-resolution above, this is possible because there is clearly a lower ``cutoff'' in the mass functions (they are not divergent to small mass), but requires a mass resolution of $\sim0.01-0.1\,\mu^{2}\,M_{d}$ (depending on the exact parameters). This is equivalent to a spatial resolution of $\epsilon_{r}\sim0.02-0.2\,Q^{1/2}\,h$, i.e.\ a small fraction of the disk scale-height $h$. This also agrees quite well with the spatial/numerical resolution where (at fixed time evolution) many of the studies above begin to see some convergence \citep[e.g.][]{meru:2012.nonconvergence.disk.frag,rice:2012.convergence.disk.frag}, but it is an extremely demanding criterion.

%\vspace{-0.5cm}
\subsection{The Magneto-Rotational Regime}
\label{sec:mri}

In the regime where the disk is magnetized and ionized, the magneto-rotational instability (MRI) can develop, driving turbulence even if the cooling rate is low and $Q\gg1$. We therefore next consider the simple case where there is {\em no} gravo-turbulent instability ($t_{\rm cool}\rightarrow \infty$), but the MRI is present.

%\vspace{-0.5cm}
\subsubsection{General Scalings}
\label{sec:mri:general}

Given MRI and no other driver of turbulence, Alfv{\'e}n waves will drive turbulence in the gas to a similar power spectrum to the hydrodynamic case (within the range we examine where the power spectrum shape makes little difference), with driving-scale rms $\langle v_{t}^{2} \rangle^{1/2} \approx v_{\rm A}$. In terms of the traditional $\mbeta$ parameter (ratio of thermal pressure to magnetic energy density; $\mbeta\rightarrow0$ as magnetic field strengths increase), $\mbeta = 2\,c_{s}^{2}/v_{\rm A}^{2}$, so the rms driving-scale Mach number is $\Mach\approx\sqrt{2\,\mbeta^{-1}}$. Magnetically-driven turbulence is close to purely solenoidal, so $b\approx1/3$ and $\Machcompressive=b\,\Mach\approx (\sqrt{2}/3)\,\mbeta^{-1/2}$. 

We stress, though, that what is important is the {\em saturated} local plasma $\mbeta=\mbeta_{\rm sat}$, which can be very different from the initial mean field $\mbeta_{0}$ threading the disk. As the MRI develops, the plasma field strength increases until it saturates in the fully nonlinear mode. Direct simulations have shown that for initial fields $\mbeta_{0}\lesssim10^{4}$, $\mbeta_{\rm sat} \sim 1/3-2/3$ \citep[see e.g.][and references therein]{bai:2012.mri.saturation.turbulence,fromang:2012.mri.turbulence.outflows}. The saturation occurs in rough equipartition with the thermal and kinetic energy densities -- i.e.\ the turbulence is trans-sonic or even super-sonic ($\Machcompressive \sim \sqrt{2/3} \sim 0.8$). In the weak-field limit, however, with $\mbeta_{0}\gtrsim 10^{4}$, the saturation is much weaker, with $\mbeta_{\rm sat}\sim10-20$, so $\Machcompressive\sim0.1$.

These same simulations allow us to directly check our simple scaling with $v_{\rm A}$; the authors directly measure the rms standard deviation in the (linear) density $\delta = \rho/\langle\rho\rangle$, which for a lognormal density distribution is (by our definitions) identical to $\Machcompressive$. For four simulations with $\mbeta_{0} = 10^{2}$, $10^{3}$, $10^{3}$ (but higher-resolution), $10^{4}$, they see (midplane) saturation $\mbeta_{\rm sat} = 0.4,\,1.1,\,0.7,\,18$ and $\langle\rho^{2}\rangle^{1/2}/\langle\rho\rangle = 0.60,\,0.43,\,0.53,\,0.13$ (compared to a predicted $\langle\rho^{2}\rangle^{1/2}/\langle\rho\rangle = \Machcompressive = (\sqrt{2}/3)\,\mbeta_{\rm sat}^{-1/2} = 0.72,\,0.44,\,0.55,\,0.11$, respectively). Moreover, these and a number of additional simulations have explicitly confirmed that our lognormal assumption (in the isothermal case) remains a good approximation for the shape of the density PDF \citep[see][]{kowal:2007.log.density.turb.spectra,lemaster:2009.density.pdf.turb.review,kritsuk:2011.mhd.turb.comparison,molina:2012.mhd.mach.dispersion.relation}. So for a given $\Machcompressive$ and $Q$, our previous derivations remain valid.

%\vspace{-0.5cm}
\subsubsection{A Statistical Stability Criterion}
\label{sec:mri:stability}

The strong-field limit therefore leads to large density fluctuations. However, strong magnetic fields will also provide support against gravity, modifying the collapse criterion; this appears in Eq.~\ref{eqn:sigmagas}. But for a given $\mbeta$, this simply amounts (to lowest order) to the replacement $c_{s}^{2}\rightarrow c_{s}^{2} + v_{A}^{2} = c_{s}^{2}\,(1 + 2\,\mbeta^{-1})$. Because $Q\propto c_{s}$, near the Toomre scale, this is approximately equivalent to raising the stability parameter as $Q\rightarrow Q_{\rm eff} \equiv Q(v_{\rm A}=0)\,\sqrt{1 + 2\,\mbeta^{-1}}$ (where $Q(v_{\rm A}=0)$ is the $Q$ including only thermal support). The energy and momentum of the bulk flows in the gas turbulence also provides support against collapse, so the ``effective dispersion'' in Eq.~\ref{eqn:sigmagas} includes all three effects; however this is already explicitly accounted for in our previous calculations for any $\Mach$. But while the effective $Q_{\rm eff}$ increases in the strong-field limit with $\mbeta^{-1/2}$, so does $\Machcompressive$, and the $Q$ needed for statistical stability on long timescales (Fig.~\ref{fig:Qmin}) increases {exponentially} with $\Machcompressive$ -- so the net effect of MRI is always to {\em increase} the probability of stochastic collapse.

Putting this into our general criterion Eq.~\ref{eqn:Qmin}, we can write the statistical stability requirement
\begin{align}
\nonumber Q(v_{\rm A}=0) &\equiv \frac{c_{s}\,\kappa}{\pi\,G\,\Sigma_{\rm gas}} 
\approx \frac{c_{s}\,\Omega}{\pi\,G\,\Sigma_{\rm gas}} 
\\
\nonumber & \ge \frac{1}{2\,\sqrt{1+2\,\mbeta_{\rm sat}^{-1}}}\,\exp{[\sqrt{2\,\ln{(\tau_{0})}\,\ln{(1+2/9\,\mbeta_{\rm sat}^{-1})}}]} \\
\label{eqn:Qmin.mri}
 &\gtrsim \frac{1}{2\,\sqrt{1+2\,\mbeta_{\rm sat}^{-1}}}\,\exp{[(2/3)\,\mbeta_{\rm sat}^{-1/2}\,\sqrt{\ln{\tau_{0}}}]}
\end{align}
where $\tau_{0}\equiv t_{0}\,\mu^{-2}\,\Omega$ as before and the latter uses the fact that $\beta$ is not extremely small in the cases of interest. 
Integrated over the lifetime of the disk, this becomes
\begin{align}
\frac{c_{s}\,\Omega}{\pi\,G\,\Sigma_{\rm gas}} 
& \gtrsim  \frac{1}{2\,\sqrt{1+2\,\mbeta_{\rm sat}^{-1}}}\,\exp{(2.9\,\mbeta_{\rm sat}^{-1/2})}
\end{align}
This increases rapidly with increasing magnetic field strength: $Q_{\rm min}(v_{\rm A}=0) \approx 15,\,7,\,3,\,1.2$ for $\mbeta_{\rm sat} = 1/3, 2/3, 2, 10$. 

So MRI with saturation $\mbeta_{\rm sat}\lesssim 10$ (``seed'' $\mbeta_{0} \lesssim 10^{4}$) will make even $Q>1$ disks statistically unstable, without the need for any other source of turbulence. On the other hand, weak-field MRI with $\mbeta_{\rm sat}\gtrsim 10$ produces only small corrections to statistical stability.

%\vspace{-0.5cm}
\subsection{Convective Disks}
\label{sec:convection}

A number of calculations have also shown that proto-planetary disks are convectively unstable over a range of radii \citep[][]{boss:2004.convective.cooling.direct.collapse,boley:2006.protoplanetary.disk.w.cooling,mayer:2007.rt.sim.dircoll.planets}. Most simulations which see convection have also seen fragmentation, which has been interpreted as a consequence of convection enhancing the cooling rates until they satisfy the \citet{gammie:2001.cooling.in.keplerian.disks} criterion for fragmentation. But \citet{rafikov:2007.convect.cooling.grav.instab.planets} and others \citep{cai:2006.disk.instab.vs.metallicity,cai:2008.protoplanet.disk.w.rad} have argued that while convection can and should develop in these circumstances, the radiative timescales at the photosphere push the cooling time above this threshold. However, as we have discussed above, that would not rule out rarer, stochastic direct collapse events. 

Consider a polytropic thin disk; this is convectively unstable when it satisfies the Schwarzschild criterion
\be
\frac{{\rm d}\ln{T}}{{\rm d}\ln{P}} > \frac{\gamma-1}{\gamma}
\ee
Following \citet{lin.papaloizou.1980:mmsn}, \citet{bell:1994.protostellar.disk.model}, and \citet{rafikov:2007.convect.cooling.grav.instab.planets}, using the fact that disk opacities can be approximated by $\kappa\approx \kappa_{0}\,P^{\alpha}\,T^{\beta}$ ($P$ the midplane pressure), this can also be writted $(1+\alpha)/(4-\beta) > (\gamma-1)/\gamma$. For the appropriate physical values, this implies strong convective instability in disks with $T\lesssim 150\,K$ (where $\kappa$ is dominated by ice grains) %; $\alpha=0$, $\beta=2$, $\gamma=7/5$), 
and at higher temperatures $\gtrsim1.5\times10^{3}\,$K when grains sublimate, and marginal convective instability in between. So this should be a common process.

A convective disk can then accelerate gas via buoyancy at a rate comparable to the gravitational acceleration, implying mach numbers $\Mach^{2}\approx0.25-1$ (depending on the driving gradients; recall also this is the three-dimensional $\Mach$) at the scale height where the disk becomes optically thin.\footnote{From mixing-length theory, we can equate the convective energy flux at the scale-height $F_{\rm conv}=2\,\rho\,C_{p}\,T\,v^{3}/h\,g_{\rm grav}$ (where at $\sim h$, the acceleration $g_{\rm grav}\approx\Omega^{2}\,h$, $h\approx c_{s}/\Omega$, and for the relevant parameters $C_{p}\approx1.25\times10^{8}$ in cgs) to the cooling flux $\sigma_{\rm SB}\,T_{\rm eff}^{4}$. This gives us the approximate estimate $\Mach \approx 0.5\,(T_{\rm eff}/200\,{\rm K})\,(\Sigma_{\rm gas}/1000\,{\rm g\,cm^{-2}})^{-1/3}\,(\Omega^{-1}/{\rm yr})^{1/3}$. This agrees well with the simulations in the text when $\Mach<1$, but extrapolates to super-sonic values at low $\Sigma$ and/or large $r_{\ast}$, so convection could be considerably more important than we estimate if it does not saturate at velocities $\sim c_{s}$.} Buoyancy-driven turbulence is primarily solenoidal forcing, so $b\approx1/3$ while $\Mach\sim0.5 - 1$, leading to a ``maximal'' $\Machcompressive\sim 0.2 - 0.3$ (assuming the convection cannot become supersonic; this is approximately what is measured in these simulations). If this saturation level is independent of $Q$ (provided the disk is convectively unstable at all), we then simply need to examine Fig.~\ref{fig:Qmin} to determine $Q_{\rm min}$ for statistical stability; from Eq.~\ref{eqn:Qmin} this is approximately
\be
\label{eqn:Qmin.convection}
Q \ge 0.5\,\exp{[\Machcompressive\,\sqrt{2\ln{\tau_{0}}}]} \sim 3
\ee
Thus, while this does not dramatically alter the behavior of the stability criterion $Q$, it does systematically increase the threshold $Q$ for statistical stability by a non-trivial factor. And indeed, in the simulations of \citet{mayer:2007.rt.sim.dircoll.planets} and \citet{boss:2004.convective.cooling.direct.collapse}, fragmentation occurs when convection is present at radii where $Q\approx1.4-1.8 > 1$.

%\vspace{-0.5cm}
\subsection{Additional Sources of Turbulence}
\label{sec:misc.turb}

There are many additional processes that may drive turbulence in proto-planetary and other Keplerian disks, but under most regimes they are less significant for our calculation here.

In the midplane of a protoplanetary disk, where large grains and boulders settle and are only weakly aerodynamically coupled to the gas, Kelvin-Helmholtz and streaming instabilities generate turbulence. However these only operate in a thin dust layer, and appear to drive rather small Mach numbers in the gas, so are unlikely to be relevant for direct collapse in the gas and we do not consider them further \citep[see e.g.][]{bai:2010.streaming.instability,shi:2012.streaming.instability}. They may, however, be critical for self-gravity of those grains participating in the instabilities themselves -- a more detailed investigation of this possibility is outside the scope of this work (since our derivation does not apply to a weakly coupled, nearly-collisionless grain population), but extremely interesting for future study.

Radiative instabilities should also operate if the disk is supported by radiation pressure, and/or in the surface layer if a wind is being radiatively accelerated off the disk by central illumination. The former case is not expected in the physical disk parameter space we consider; but if it were so, convective, magneto-rotational, and photon-bubble instabilities are also likely to be present \citep[][]{blaes:2001.radiation.disk.instabilities,thompson:2008.grav.instab.in.rad.pressure}, which will drive turbulence that saturates in equipartition between magnetic, radiation, and turbulent energy densities, i.e.\ produce the equivalent of $\Mach\sim1$ throughout the disk (giving results broadly similar to the strong-field MRI case). 
In the wind case, the Mach numbers involved can be quite large (since material is accelerated to the escape velocity), but unless the surface layer includes a large fraction of the mass, it is unlikely to be important to the process of direct collapse. 

In the case of an AGN accretion disk, local feedback from stars in the disk may also drive turbulence (as it does in galactic disks), and this can certainly be significant in the outer regions of the disk where star formation occurs \citep[see][]{thompson:rad.pressure}. In that case the turbulence may even be super-sonic, in which case a more appropriate model is that developed in \paperone-\papertwo. In the inner parts of the disk, though, where the turbulence is sub-sonic, we are not necessarily interested in rare single star-formation events.

\begin{figure}
    \centering
    \plotonesize{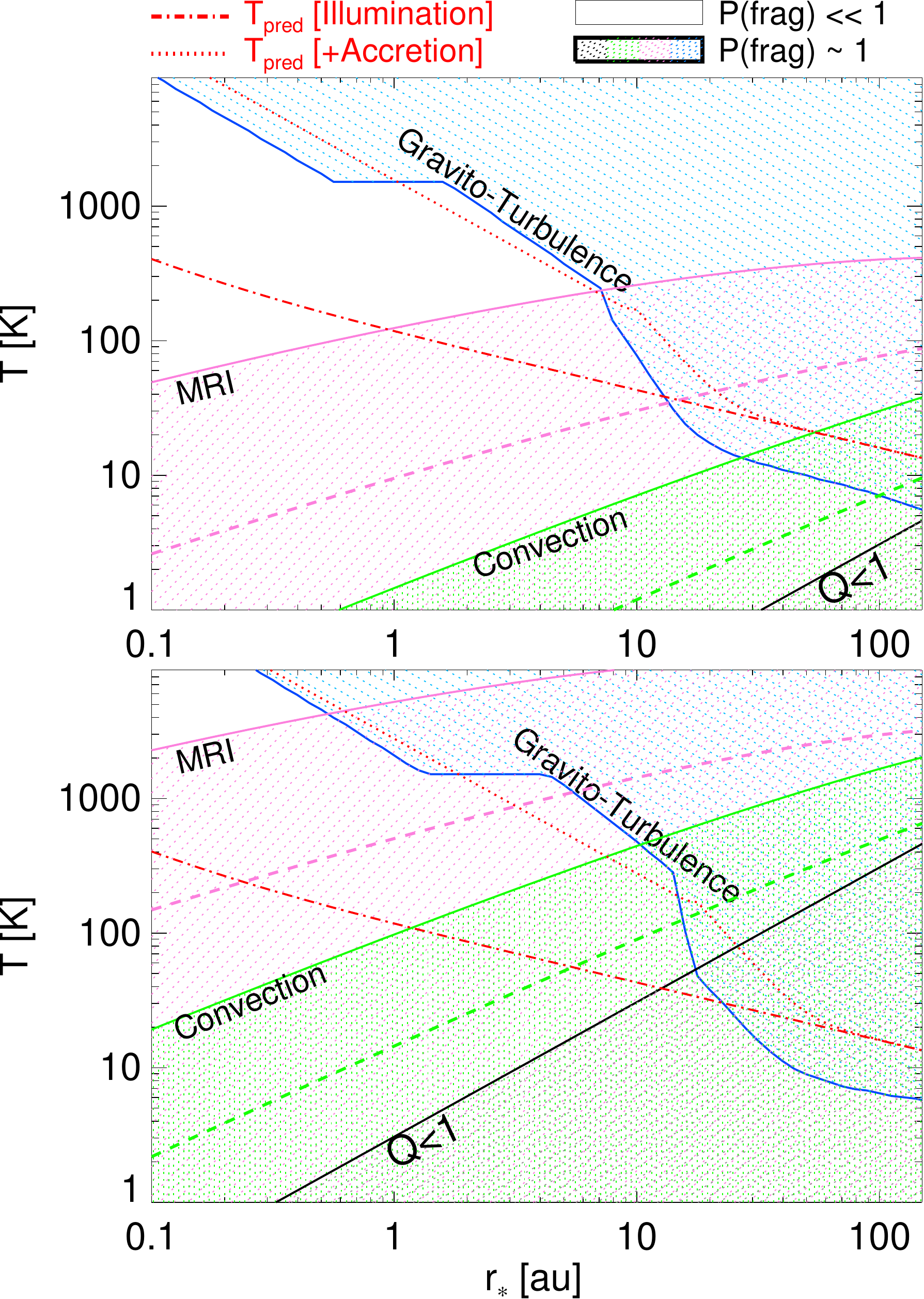}{0.99}
    \caption{Shaded regions show the range of disk temperatures, at a given radius around a solar-type star, in which a proto-planetary disk is statistically unstable (i.e.\ has an order-unity probability $\Pfragint\sim1$ of at least one ``fragmentation'' or direct collapse event in a timescale $\approx 1\,$Myr). 
    {\em Top:} Disk with $\Sigma=1000\,{\rm g\,cm^{-3}}\,(r_{\ast}/{\rm au})^{-1}$. 
    {\em Bottom:} $10\,x$ higher $\Sigma$ ($\Sigma_{0,\,1000}=10^{4}\,{\rm g\,cm^{-2}}$; $M_{\rm disk}(<10\,{\rm au})\sim0.05\,M_{\ast}$). 
    Black is the standard Toomre $Q<1$ (catastrophic fragmentation). 
    Other shaded regions correspond to different mechanisms driving turbulence, with $\Machcompressive$, $Q$, etc.\ calculated self-consistently for $\Sigma(r_{\ast})$, $T$, and $M_{\ast}$ (see \S~\ref{sec:specific}).
    Green: Temperature where $\Pfragint\sim1$ if the disk is convectively unstable (\S~\ref{sec:convection}; Eq.~\ref{eqn:Qmin.convection}). Solid/dashed lines correspond to the higher/lower $\Mach$ estimated from convective driving in simulations.
    Pink: Temperature where $\Pfragint\sim1$ if the disk has a saturated (strong-field) magneto-rotational instability (MRI; \S~\ref{sec:mri}; Eqn~\ref{eqn:Qmin.mri}); again solid/dashed correspond to stronger/weaker limits on saturation ($\mbeta_{\rm sat} = 0.45 - 0.85$, from seed $\mbeta_{0}=10^{2}-10^{3}$). 
    Blue: Gravito-turbulence (\S~\ref{sec:beta.disks}; Eqn.~\ref{eqn:Qmin.gravitoturb}); here, {\em higher} $T$ corresponds to faster cooling, hence higher $\Machcompressive$ (Eqn~\ref{eqn:mach.alpha}), and increased $\Pfragint$.
    Red lines show the calculated $T$ for a disk with the given $\Sigma(r_{\ast})$ from illumination by a solar-type star (dot-dashed) \&\ illumination plus accretion with $\dot{M}=3\times10^{-7}\,\msun\,{\rm yr^{-1}}$ (dotted).
    Even in MMSN, radii $\gtrsim$ a few au are statistically unstable via gravito-turbulence; smaller radii are statistically unstable for $\dot{M}\gtrsim 3\times10^{-7}\,\msun\,{\rm yr^{-1}}$. Strong-field MRI is also capable of generating sufficient fluctuations for direct collapse down to $\sim0.1-1\,$au. 
    \label{fig:T.vs.r}}
\end{figure}

\begin{figure}
    \centering
    \plotonesize{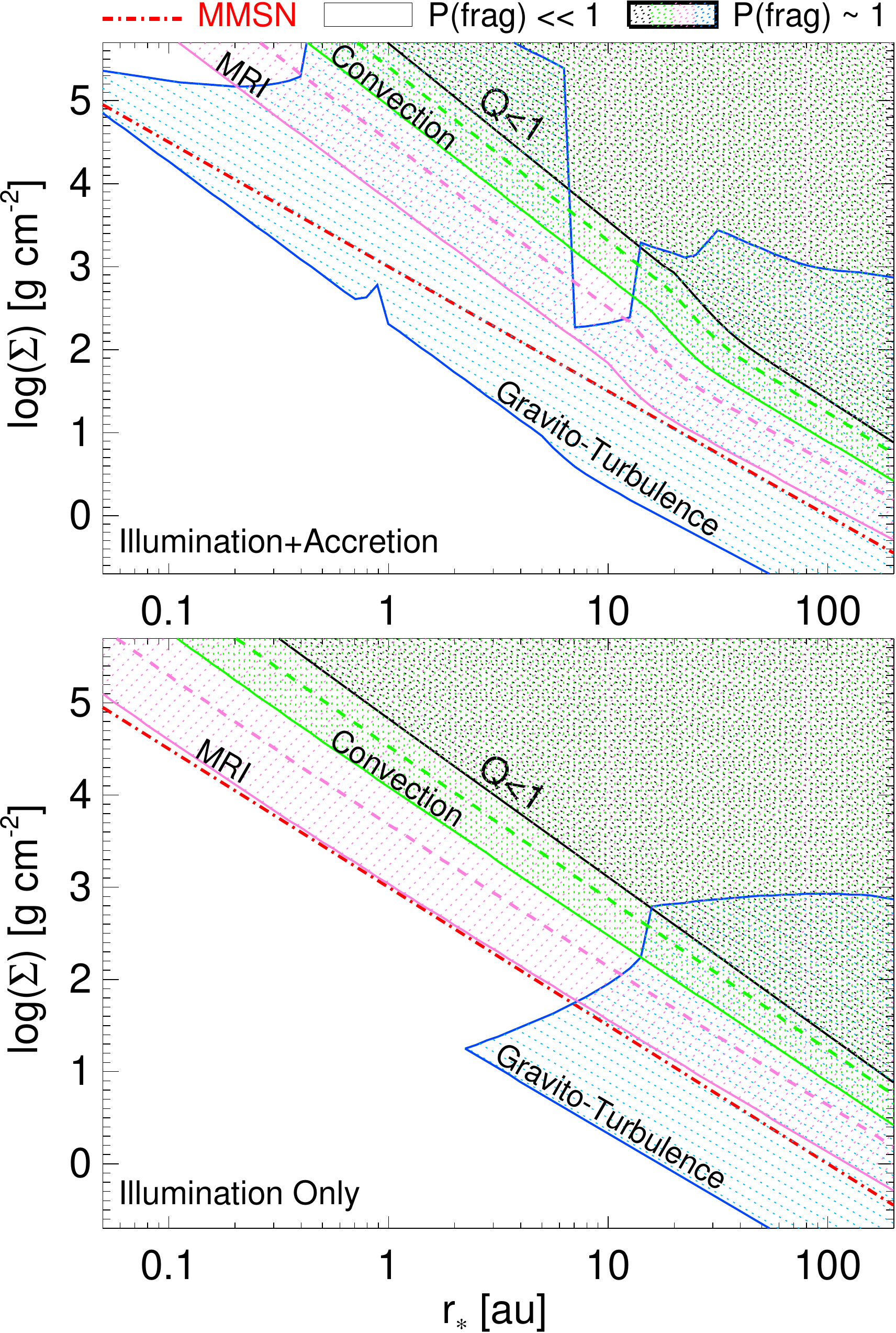}{0.99}
    \caption{Shaded regions show the range of surface densities which are statistically unstable (have an order-unity probability $\Pfragint\sim1$ of a direct collapse event in a timescale $\approx1\,$Myr, as Fig.~\ref{fig:T.vs.r}), in a proto-planetary disk at a given radius around a solar-type star. For each $\Sigma(r_{\ast})$, we calculate $T$ self-consistently including illumination and accretion with $\dot{M}=3\times10^{-7}\,\msun\,{\rm yr^{-1}}$ ({\em top}) or illumination only ({\em bottom}); see \S~\ref{sec:example.model}. Each shaded range corresponds to a different candidate source of turbulent density fluctuations, as in Fig.~\ref{fig:T.vs.r}. Red line shows the MMSN, for comparison. Strong MRI can promote collapse in a MMSN at all radii; gravito-turbulence in even lower-density disks at $\gtrsim$\,au, and smaller radii if $\dot{M}\gtrsim3\times10^{-7}\,\msun\,{\rm yr^{-1}}$. If these are not active, convection can promote collapse in higher-density disks with $\Sigma\gtrsim10\,\Sigma_{\rm MMSN}$. 
    \label{fig:sigma.vs.r}}
\end{figure}

%\vspace{-0.5cm}
\section{Example: ProtoPlanetary Disk and Fragmentation Radii}
\label{sec:example.model}

We now apply the statistical stability criteria derived above to a specific model of a proto-planetary disk. This is highly simplified, but it allows us to estimate physically reasonable sound speeds, cooling times, and other parameters, so allows us to ask whether our revised statistical stability criteria are, in practice, important. 

%\vspace{-0.5cm}
\subsection{Disk Model Parameters}
\label{sec:example.model:parameters}

For convenience, consider a disk with a simple power-law surface density profile
\be
\Sigma = \Sigma_{0,\,1000}\,{\rm 1000\,{\rm g\,cm^{-2}}}\,(r_{\ast}/{\rm au})^{-\alpha}
\ee
The minimum mass solar nebula (MMSN) corresponds to $\Sigma_{0,\,1000}\sim1$ and $\alpha\approx1.5$, but we consider a range in these parameters below.

For a passive flared disk irradiated by a central star with radius $=R_{\ast}$ and temperature $=T_{\ast}$, the effective temperature is \citep{chiang:1997.protostellar.disk.sed}:
\be
T_{\rm eff,\,\ast} \approx {\Bigl(} \frac{\alpha_{T}\,R_{\ast}^{2}}{4\,r_{\ast}^{2}} {\Bigr)}^{1/4}\,T_{\ast} 
\ee
where for a solar-type star $\alpha_{T} \approx 0.005\,(r_{\ast}/{\rm au})^{-1} + 0.05\,(r_{\ast}/{\rm au})^{2/7}$, $R_{\ast}=R_{\sun}$ and $T_{\ast}=6000\,K$. If, instead of irradiation, disk heating is dominated by energy from steady-state accretion with some $\dot{M}$, energy balance requires an effective temperature (i.e.\ disk flux) 
\be
T_{\rm eff,\,acc} \approx {\Bigl[}\frac{3}{8\pi}\,\frac{\dot{M}\,\Omega^{2}}{\sigma_{\rm SB}} {\Bigr]}^{1/4}
\ee
where $\sigma_{\rm SB}$ is the Stefan-Boltzmann constant. The temperature of interest for our purposes, however, is the midplane temperature $T_{\rm mid}$, since this is where the disk densities are largest and what provides the $c_{s}$ resisting collapse; this is related to $T_{\rm eff}$ by a function of opacity and $\Sigma$ which we detail in Appendix~\ref{sec:appendix:tempcalc}. But having determined $T_{\rm eff}$ and $\Sigma$, it is straightforward to calculate $T_{\rm mid}$; the sound speed is $c_{s} = \sqrt{k_{B}\,T_{\rm mid}/\mu}$ where $k_{B}$ is the Boltzmann constant and $\mu$ is the mean molecular weight.

This is sufficient to specify most of the parameters of interest. In the gravito-turbulent model, however, we also require an estimate of the cooling time to estimate $\cbeta\equiv t_{\rm cool}\Omega$. \citet{rafikov:2007.convect.cooling.grav.instab.planets} calculate the approximate cooling time for a convective and radiative disk (depending on whether or not it is convective and, if so, accounting for the rate-limiting of cooling by the disk photosphere). This gives 
\begin{align}
t_{\rm cool} &= \frac{\Sigma\,c_{s}^{2}}{\sigma_{\rm SB}\,T^{4}}\,f(\tau) \\
\nonumber &\approx 2\times10^{4}\,{\rm yr}\,{\Bigl(}\frac{\Sigma}{10^{3}\,{\rm g\,cm^{-2}}}{\Bigr)}\,{\Bigl(} \frac{T}{100\,{\rm K}} {\Bigr)}^{-3}\,\frac{f(\tau)}{10^{3}}
\end{align}
where $f(\tau)$ is a function (shown in Appendix~\ref{sec:appendix:tempcalc}) of the opacity which interpolates between the optically thin/thick, and convective/radiative regimes. 

With these parameters calculated, for a given assumption about what drives the turbulence -- e.g.\ MRI, gravitoturbulence, convection, etc. -- the compressive Mach number $\Machcompressive$ can be calculated following \S~\ref{sec:specific}. We also technically need to assume the details of the turbulent spectral shape, for which we will assume a spectral index $p=2$ and non-intermittent $T=0$, as well as the gas equation of state, for which we take $\gamma=7/5$, appropriate for molecular hydrogen. But these choices have small effects on our results, as shown in Fig.~\ref{fig:mfs}.

%\vspace{-0.5cm}
\subsection{The Characteristic Initial Fragment Mass}
\label{sec:example.model:mass}

In Fig.~\ref{fig:mass.vs.r}, we use this model to calculate the expected mass of a self-gravitating ``fragment.'' Varying $\Sigma_{0,\,1000}$, $\alpha$, and $M_{\ast}$, we calculate the expected $T$ and $Q$, assuming a constant accretion rate of $\dot{M} = 3\times10^{-7}\,\msun\,{\rm yr^{-1}}$ at all radii,\footnote{This may not be self-consistent, since $\dot{M}$ could vary with disk parameters, but there is no straightforward {\em a priori} expectation for $\dot{M}$, and we only intend this as a guide, in any case.} which dominates the disk temperature inside $\sim10$\,au. Given this, we calculate the mass spectrum as Fig.~\ref{fig:mfs}, and define an average $\langle M_{\rm collapse} \rangle$ (the mass-weighted, spectrum integrated mass). Technically this depends on $\Machcompressive$ hence the turbulent driving mechanism, but the dependence is weak so we just assume $\Machcompressive=0.1$ in all cases. 

In each case, $\langle M_{\rm collapse}\rangle\sim \mu^{2}\,Q^{2}\,M_{\rm disk}(<r_{\ast})$ as expected. Since $\mu\equiv (\pi\,\Sigma\,r_{\ast}^{2})/M_{\ast}$, this increases with disk surface density or mass, and also with increasing disk-to-stellar mass ratio. Recall $T_{\rm eff,\,acc}\propto (\dot{M}\,\Omega^{2})^{1/4} \propto r_{\ast}^{-3/4}$ and $T_{\rm eff,\,\ast}\propto r_{\ast}^{-1/2}$. So modulo opacity corrections we expect $\langle M_{\rm collapse} \rangle \propto r_{\ast}^{3\,(1.5-\alpha)}$ at small radii $\lesssim$ a few au, weakly increasing with $r_{\ast}$; and $\langle M_{\rm collapse} \rangle \propto r_{\ast}^{0.5+3\,(1.5-\alpha)}$ at large $r_{\ast}$, increasing more rapidly. 

This is only the initial self-gravitating, bound mass -- it may easily evolve in time, as discussed in \S~\ref{sec:discussion}. However, it is interesting that there is a broad range of masses possible, with Earth and super Earth-like masses more common at $\lesssim1\,$au and giant planet masses more common at $\gtrsim10$\,au.

%\vspace{-0.5cm}
\subsection{Disk Temperatures at Which Direct Collapse Occurs}
\label{sec:example.model:temperature}

Given a mass profile, then for a source of turbulence in \S~\ref{sec:specific} we can translate the criteria  for statistical stability -- a probability $\Pfragint\ll1$ of forming a fragment in a characteristic timescale $\sim$Myr -- into a range of midplane temperatures $T_{\rm mid}$. 

Fig.~\ref{fig:T.vs.r} shows this for a disk with $\Sigma_{0,\,1000}=1$ and $\alpha=1$, around a solar-type star, as well as a disk with $\Sigma_{0,\,1000}=10$. Choosing $\alpha=1.5$ gives a similar result but with the curve slopes systematically shifted. Together this spans a range in disk mass at $\sim10$\,au of $\sim0.004 - 0.07\,M_{\ast}$. We compare the expected $T_{\rm mid}(r_{\ast},\,\Sigma)$, for accretion rates $\dot{M}=0$ and $\dot{M}=3\times10^{-7}\,\msun\,{\rm yr^{-1}}$. 

There is some $T(r_{\ast})$ below which $Q<1$, so the disk will catastrophically fragment. But this is quite restrictive: $Q<1$ only when $\Sigma$ and $r_{\ast}$ are large (for $\Sigma_{0,\,1000}\sim10-100$, $r_{\ast}\gtrsim 10-100\,$au for $\alpha\sim1-1.5$, respectively; roughly where $M_{\rm disk}(<r_{\ast})\gtrsim 0.1\,M_{\ast}$). 
%at $R\gtrsim 15\,$au when $\Sigma_{0,\,1000}=10$ and $\alpha=1$; or $R\gtrsim100\,$au when $\Sigma_{0,\,1000}\sim10-100$ for $\alpha=1.5$, 

If the disk is convectively unstable, the resulting Mach numbers lead to a large temperature range over which $Q>1$, so the disk is classically stable, but $\Pfragint\sim1$; this is less likely to be relevant for a MMSN ($r_{\ast}\gtrsim50\,$au for $\Sigma_{0,\,1000}=1$) but can be sufficient at $r_{\ast}\gtrsim2-10$\,au for $\Sigma_{0,\,1000}\gtrsim10$. As discussed in \S~\ref{sec:convection}, the convective Mach numbers are somewhat uncertain, so we show the calculation for the range therein. 

Strong-field MRI -- {if/where it is active} -- produces even larger $\Mach$; this can greatly expand the range where $\Pfragint\sim1$ even in a MMSN. Again there is a range of possible $\Machcompressive$ in the saturated state, shown here. In the weak-field regime, $\Machcompressive$ is smaller, giving results very similar to the convection prediction.

Gravito-turbulence (again, if/where it is active), interestingly, has the opposite dependence on temperature. Because cooling rates grow rapidly at higher $T_{\rm mid}$, the expected $\Machcompressive$ is larger and $\Pfragint$ is {\em larger} at higher $T$ (despite higher $Q$). It is also much less sensitive to the disk surface density. If the process operates, this is sufficient to produce $\Pfragint\sim1$ at $r_{\ast}\gtrsim2-5\,$au, regardless of $\dot{M}$ for nearly all reasonable temperatures, and even at $r_{\ast}\lesssim1\,$au if there is a modest $\dot{M}$ to raise the temperature (hence cooling rate) to $\gtrsim300-1000\,$K. For the disk surface densities here, $\Pfragint\ll1$ at $r_{\ast}\lesssim1\,$au requires $\dot{M} \lesssim 3\times10^{-7}\,\msun\,{\rm yr^{-1}}$. However, we caution that at sufficiently high $T$ and $Q$, the mechanism may not operate at all.

Beyond a certain radius (which depends on which combination of these mechanisms are active), perhaps the most important result is that there may be {\em no} temperature where $\Pfragint\ll1$, for a given surface density.

%\vspace{-0.5cm}
\subsection{Surface Densities at Which Direct Collapse Occurs}
\label{sec:example.model:sigma}

In Fig.~\ref{fig:sigma.vs.r}, we calculate the surface densities (as a function of radius $r_{\ast}$ around a solar-type star) where disks are statistically unstable ($\Pfragint\sim1$) if/when different turbulent driving mechanisms are active, as in Fig.~\ref{fig:T.vs.r}. Whereas in Fig.~\ref{fig:T.vs.r} we allowed the temperature to be free, here we assume it follows our best estimate $T_{\rm mid}(r_{\ast},\,\Sigma)$ for either an accretion rate $\dot{M}=0$ or $\dot{M}=3\times10^{-7}\,\msun\,{\rm yr^{-1}}$, but freely vary $\Sigma(r_{\ast})$. 

Roughly speaking, convection and MRI systematically lower the density at all radii where $\Pfragint\sim1$. Classical instability ($Q<1$) requires $\Sigma\gtrsim30\,\Sigma_{\rm MMSN}$. If the disk is convective it can have $Q>1$ but be statistically unstable (vulnerable to direct fragmentation via turbulence density fluctuations) for $\Sigma\gtrsim10\,\Sigma_{\rm MMSN}$ at $r_{\ast}\gtrsim10\,$au; at smaller radii the threshold is sensitive to accretion-heating raising $Q$. If strong-field MRI is active, the threshold surface density for statistical instability is much lower: for small accretion rates and/or large radii $\gtrsim2-5\,$au, even the MMSN can have $\Pfragint\sim1$. Gravito-turbulence, if active, generates sufficient turbulence for statistical instability ($\Pfragint\sim1$) even at $\Sigma \sim 0.1\,\Sigma_{\rm MMSN}$, at radii $r_{\ast}\gtrsim1\,$au regardless of accretion rate, and $r_{\ast}\lesssim1\,$au for $\dot{M}\gtrsim 3\times10^{-7}\,\msun\,{\rm yr^{-1}}$.

%\vspace{-0.5cm}
\section{Discussion \&\ Conclusions}
\label{sec:discussion}

Traditionally, a disk with Toomre $Q>1$ is classically stable against gravitational collapse on all scales (modulo certain {global} gravitational instabilities). However, we show here that this is no longer true in a turbulent disk. Random turbulent density fluctuations can produce locally self-gravitating regions that will then collapse, even in $Q>1$ disks.\footnote{{Recall these arise from the super-position of many smaller perturbations/turbulent structures, not necessarily a ``global'' forcing event.
}} Formally, the probability of such an event is always non-zero, so strictly speaking turbulent disks are never ``completely'' stable, but can only be so {\em statistically}, if the probability of forming a self-gravitating fluctuation is small over the timescale of interest. Moreover, we can analytically predict the probability, as a function of total self-gravitating mass, of the formation of such a region per unit time in a disk (or disk annulus) with given properties.

We do this using the excursion-set formalism developed in \paperone-\papertwo, which allows us to use the power spectra of turbulence to predict the statistical properties of turbulent density fluctuations. In previous papers, this has been applied to the structure of the ISM in galactic or molecular cloud disks; however, we show it is straightforward to extend to a Keplerian, proto-planetary disk. The most important difference between the case here and a galactic disk is that in the galactic case, turbulence is highly super-sonic ($\Mach\sim10-100$), as opposed to sub/trans-sonic. And in galactic disks, cooling is rapid and the disk is globally self-gravitating (non-Keplerian), so systems almost always converge rapidly to $Q\approx1$ (see \citealt{hopkins:fb.ism.prop}); here, we expect a wider range of $Q$. And finally, proto-planetary disks are very long-lived relative to their local-dynamical times, so even quite rare events (with a probability of, say, $10^{-6}$ per dynamical time) may be expected over the disk lifetime.

At $Q\approx1$, disks with $\Machcompressive\gtrsim0.1$ are classically stable but statistically unstable: they are likely, over the lifetime of the disk, to experience at least an order-unity number of ``fragmentation'' events (formation of self-gravitating, collapsing masses). As expected, higher-$Q$ disks are ``more stable'' (although again we stress this is only a probabilistic statement); for $Q\sim3-5$, values of $\Machcompressive\gtrsim0.3-0.5$ are required for statistical instability. If the turbulence is transsonic ($\Machcompressive\sim1-3$), values as large as $Q\sim40-1000$ can be statistically unstable! Generally speaking, we show that statistical stability (i.e.\ ensuring that the probability of a stochastic direct collapse event is $\ll 1$) integrated over some timescale of interest $t_{0}$ requires a $Q_{\rm min}$ 
\begin{align}
\nonumber Q_{\rm min} &\approx 0.5\,\exp{[\sqrt{2\,\ln{(t_{0}\,\mu^{-2}\,\Omega)}\,\ln{(1+\Machcompressive^{2})}}]}  \\
\nonumber &\sim 0.5\,\exp(6\,\sqrt{\ln{(1+\Machcompressive^{2})}})
\end{align}
(Fig.~\ref{fig:Qmin} \&\ Eq.~\ref{eqn:Qmin}). At the Mach numbers of interest, this is exponentially increasing with $\Mach$! 

This is a radical revision to traditional stability criteria. However, the traditional Toomre $Q$ criterion is not irrelevant. It is a necessary, but not {\em sufficient}, criterion for statistical stability, which should not be surprising since its derivation assumes a homogenous, non-turbulent disk. If $Q\ll1$, then ``catastrophic'' fragmentation occurs even for $\Machcompressive\rightarrow0$; all mass in the disk is (classically) unstable to self-gravity and collapse proceeds on a single free-fall time. If $Q>1$, fragmentation transitions to the stochastic (and slower) statistical mode calculated here, dependent on random turbulent density fluctuations forming locally self-gravitating regions.

Likewise, the criterion in \citet{gammie:2001.cooling.in.keplerian.disks}, that the cooling time be longer than a couple times the dynamical time, is not sufficient for statistical stability. For a given turbulent Mach number and $Q$, we show that assuming a stiffer equation of state has quite weak effects on the ``stochastic'' mode of fragmentation; in fact, even pure adiabatic gas (no cooling) produces very similar statistics if it can sustain a similar $\Machcompressive$. Again, the key is that the \citet{gammie:2001.cooling.in.keplerian.disks} criterion is really about the prevention of {\em catastrophic} fragmentation; as noted therein, a sufficiently slow cooling time allows turbulence to maintain a steady state $Q\sim1$, and dissipation of that turbulence (driven by gravitational density waves and inflow) can maintain the gas thermal energy ($c_{s}$). Thus faster cooling leads to catastrophic fragmentation of most of the mass on a single dynamical time ($Q<1$ and $\Machcompressive\gg1$). And indeed this is the case from \paperone\ in a galactic disk, where $t_{\rm cool}\ll t_{\rm dyn}$ and the mass is only ``recycled'' back into the diffuse medium by additional energy input (from stellar feedback). 

However, provided the \citet{gammie:2001.cooling.in.keplerian.disks} criterion is met and $Q>1$ everywhere, we still predict fragmentation in the ``stochastic'' mode if gravito-turbulence operates. The cooling time is then important insofar as it changes the equilibrium balance between turbulent and thermal energy, i.e.\ appears in $Q$ and governs $\Machcompressive\propto (t_{\rm cool}\,\Omega)^{-1/2}$. This, indeed, has now been seen in a growing number of simulations (see references in \S~\ref{sec:intro}), involving either larger volumes and/or longer runtimes. We consider the application of our theory to these specific models in \S~\ref{sec:beta.disks}; this allows us to predict a revised cooling-time criterion in this ``mode,'' required for statistical stability over any timescale of interest:
\begin{align}
\label{eqn:betamin.maintext}
\cbeta\equiv\frac{t_{\rm cool}}{\Omega^{-1}} %&\approx \frac{4}{3\,\gamma\,(\gamma-1)}\,\frac{\ln{(\tau_{0})}}{[\ln{(2\,Q)}]^{2}} \\ 
& > \frac{4}{3\,\gamma\,(\gamma-1)}\,\frac{\ln{[t_{0}\,\mu^{-2}\,\Omega)]}}{[\ln{(2\,Q)}]^{2}}
\end{align}
This provides an excellent explanation for the results in these simulations, and resolves the apparent discrepancies between them noted in \S~\ref{sec:intro}.

If another process is able to drive turbulence, then stochastic direct collapse might occur with even longer cooling times. We show that if the MRI is active and saturates at strong-field $\beta_{\rm sat}\sim1$, the required $Q$ for complete suppression of fragmentation can be very large ($\gtrsim 10-15$; scaling with $\beta_{\rm sat}$ as Eq.~\ref{eqn:Qmin.mri}), {\em independent} of the cooling time. If the MRI is not active (in the dead zone, for example), or if it saturates at weak-field values $\beta_{\rm sat}\gtrsim10$, and cooling is slower than the limit above, then convection may be the dominant driver of turbulence. This is sufficient to produce stochastic fragmentation events in the range $Q\sim1-3$, though probably not much larger. 

We apply these calculations to specific models of proto-planetary disks that attempt to self-consistently calculate their temperatures and cooling rates. Doing so, we show that the parameter space where statistical instability and stochastic fragmentation may occur is far larger than that of classical instability (where $Q<1$), and can include most of the disk even in a MMSN. Gravito-turbulence appears to be the most important channel driving stochastic fragmentation when it is active, and is sufficient to produce an order-unity number of events in disks with $\Sigma\gtrsim0.1\,\Sigma_{\rm MMSN}$ at distances $\gtrsim1\,$au (independent of accretion rate) and even at $\lesssim1\,$au (if the disk is heated by accretion rates $\gtrsim 3\times10^{-7}\,\msun\,{\rm yr^{-1}}$). At low accretion rates and/or large radii, strong-field MRI (if active) is also sufficient to drive statistical instability if $\Sigma\gtrsim \Sigma_{\rm MMSN}$. And we show that beyond a few au, the combination of gravito-turbulence, convection, and Toomre instability at low-$T$ means there may be no disk temperatures at which $\Pfragint\ll 1$ in a modest-density disk.

Ultimately, regardless of our (admittedly speculative) discussion of theoretical models for the sources of turbulence in proto-planetary disks, the key question is empirical. Are the actual Mach numbers in such disks sufficient, for their $Q$ values, to be ``interesting'' here? This remains an open question. However, \citet{hughes:2011.turb.protoplanetary.disk.obs} present some early indications of detection of turbulent linewidths in two protoplanetary disks, with inferred Mach numbers of $\sim 0.1$ and $\sim0.4$. Although uncertain, these essentially bracket the most interesting regime of our calculations here! Because of the exponential dependence of stochastic collapse on Mach numbers, future observations which include larger statistical samples and more/less massive disks, as well as constraints on whether the turbulence appears throughout the disk (since it is the midplane Mach numbers that matter most for the models here), will be critical to assess whether the processes described in this paper are expected to be relatively commonplace or extremely rare. 

We also predict the characteristic mass spectrum of fragmentation events. Sub-sonic turbulence produces a narrow mass spectrum concentrated around the Toomre mass $\sim \mu^{2}\,Q^{2}\,M_{\rm disk}$; angular momentum and shear suppress larger-scale collapse while thermal (and magnetic) pressure suppress the formation of smaller-scale density fluctuations. For typical mass ratios $\mu\sim0.1$ in the early stages of disk evolution, this corresponds reasonably well to the masses of giant planets. For smaller mass ratios $\mu\sim0.01$, which should occur at somewhat later stages of evolution, this implies direct collapse to Earth-like planet masses may be possible. Of course, such collapse will carry whatever material is mixed in the disk (i.e.\ light elements), so such a planet would presumably lose its hydrogen/helium atmosphere as it subsequently evolved (see references in \S~\ref{sec:intro}). As the turbulence approaches transsonic, the mass spectrum becomes much more broad. As shown in \papertwo, this owes to the greater dynamic range in which turbulence is important; in the limit $\Mach\rightarrow\infty$, the mass spectrum approaches a power law with equal mass at all logarithmic intervals in mass (up to the maximum disk Jeans mass). Thus within a more turbulent disk, direct collapse to a wide range of masses -- even at {\em identical} disk conditions and radii -- is expected. This may explain observed systems with a range of planet masses within narrow radii \citep[e.g.][]{carter:2012.kepler36.neighbor.giant.rocky.planets}, and it also predicts a general trend of increasing average collapse mass with distance from the central star. However, we are not accounting for subsequent orbital evolution here, and subsequent accretion (after collapse) will modify the mass spectra.

{ 
We discuss and show how these predictions change with different turbulent velocity power spectra, different gas equations-of-state, including or excluding magnetic fields, changing the disk mass profile, or allowing for (quite large) deviations from log-normal statistics in the density distributions. However, these are generally very small corrections and/or simply amount to order-unity re-normalizations of the predicted object masses, and they do not change any of our qualitative conclusions. Because of the very strong dependence of fragmentation on Mach number, the critical Mach numbers we predict as the threshold for statistical instability are insensitive to most changes in more subtle model assumptions.\footnote{Formally, allowing for correlated structure in the density field, non-linear density smoothing, different turbulent power-spectra, or intermittency and non-Gaussian statistics in the density PDF, all discussed in detail in \papertwo, within the physically plausible range, produces sub-logarithmic corrections to the $\Machcompressive$ and $Q$ criteria we derive for statistical stability.}
}

Throughout, we restrict our focus to the {\em formation} of self-gravitating regions.\footnote{{Specifically, the ``threshold'' criterion we use implies that, within the region identified, the total energy (thermal, magnetic, kinetic, plus self-gravitational) is negative; the region is linearly unstable to gravitational collapse; and the (linear, isothermal) collapse timescale ($\sim 1/\sqrt{G\,\rho_{\rm crit}}$) is shorter than each of the shear timescale ($\sim\Omega^{-1}$), the sound crossing time ($\sim R/c_{s}(\rho_{\rm crit})$), and the turbulent cascade energy or momentum ``pumping'' timescale ($\sim R/\langle v_{t}(R)\rangle$). This automatically ensures that less stringent criteria such as the Jeans, local Toomre $Q$, and Roche criteria are satisfied (see \paperone\ \&\ \papertwo\ for details).}} Following the subsequent evolution of those regions (collapse, fragmentation, migration, accretion, and possible formation of planets) requires numerical simulations to treat the full non-linear evolution \citep[see e.g.][and references therein]{kratter:2011.disk.frag.criteria,rice:2011.irradiated.disk.stability,zhu:2012.disk.frag.clump.evol,galvagni:2012.clump.collapse}. Ideally, this would be within global models that can self-consistently follow the formation of these regions. However, this is computationally extremely demanding. Even ignoring the detailed physics involved, the turbulent cascade must be properly followed (always challenging), and most important, since the fluctuations of interest can be extremely rare, very large (but still high-resolution) boxes must be simulated for many, many dynamical times. As discussed in \S~\ref{sec:intro}, this has led to debate about whether or not different simulations have (or even can be) converged. Most of the longest-duration simulations to date have been run for times $\sim1000\,\Omega^{-1}$, which for plausible $\Machcompressive$ and $Q$ may be shorter by a factor of $\sim10^{3}-10^{5}$ than the timescale on which of order a {\em single} event is expected to occur in the entire disk. But certainly in the example of planet formation, a couple of rare events are ``all that is needed,'' so this is an extremely interesting case for future study.

%\vspace{-0.7cm}
\acknowledgments 
We thank Jim Stone, Eugene Chiang, and Eliot Quataert for insightful discussions that helped inspire this paper. Support for PFH was provided by NASA through Einstein Postdoctoral Fellowship Award Number PF1-120083 issued by the Chandra X-ray Observatory Center, which is operated by the Smithsonian Astrophysical Observatory for and on behalf of the NASA under contract NAS8-03060.\\

%\vspace{-0.1cm}
\bibliography{/Users/phopkins/Documents/work/papers/ms}

\begin{appendix}
\appendixcolumns

%\vspace{-0.5cm}
\section{A.\ Details of Temperature and Cooling Rate Calculation}
\label{sec:appendix:tempcalc}

\begin{figure}
    \centering
    \plotonesize{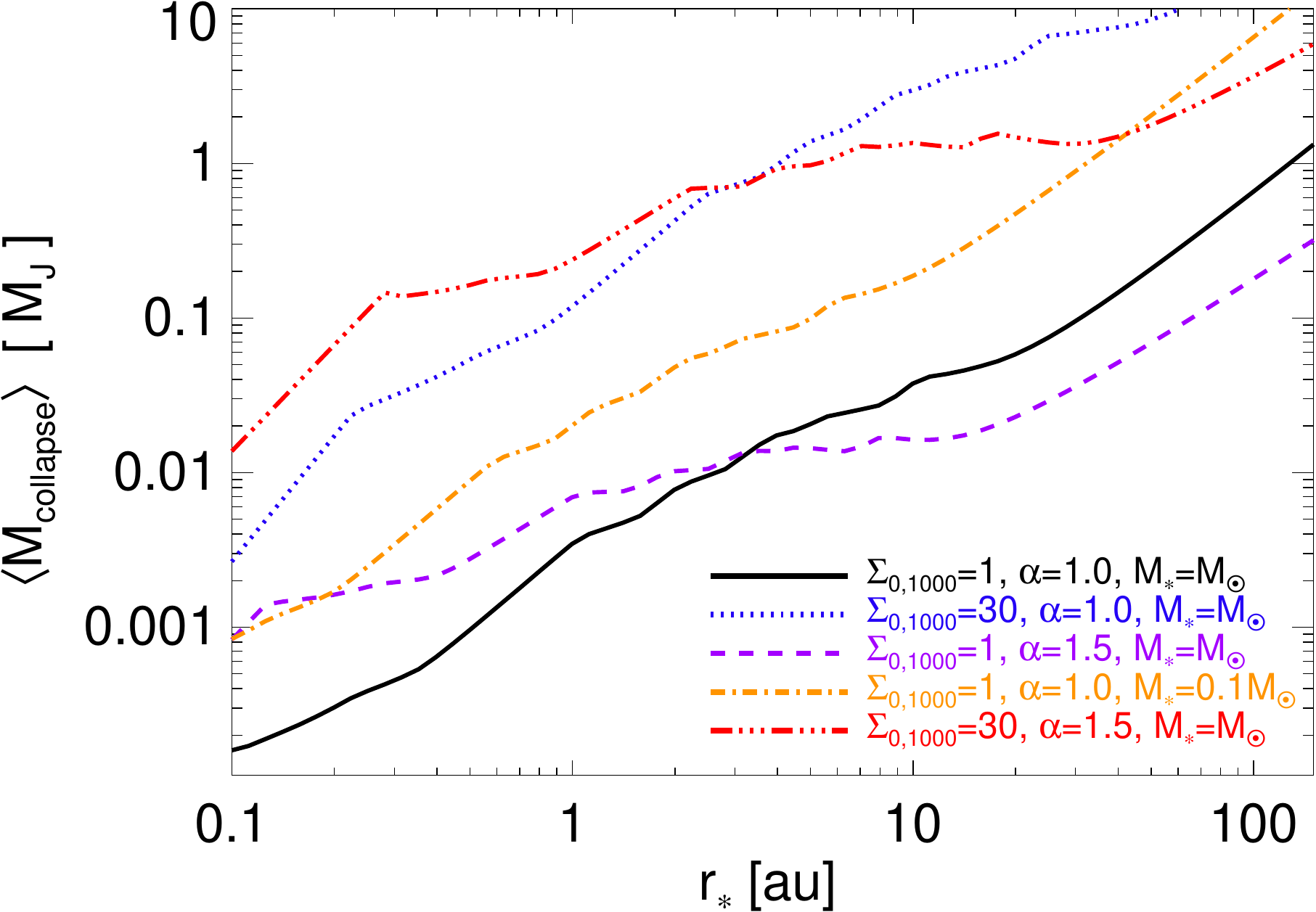}{1.01}
    \caption{Characteristic mass at collapse, as Fig.~\ref{fig:mass.vs.r}, but with a more detailed set of opacity tables and temperature calculation as described in \S~\ref{sec:appendix:tempcalc}.
    \label{fig:mass.vs.r.kappa}}
\end{figure}

Recall, from the text, in the case of disk irradiation by a central solar-type star, the effective temperature is 
\be
T_{\rm eff,\,\ast} \approx {\Bigl(} \frac{\alpha_{T}\,R_{\ast}^{2}}{4\,r_{\ast}^{2}} {\Bigr)}^{1/4}\,T_{\ast} 
\ee
\citep{chiang:1997.protostellar.disk.sed} with $\alpha_{T} \approx 0.005\,(r_{\ast}/{\rm au})^{-1} + 0.05\,(r_{\ast}/{\rm au})^{2/7}$, $R_{\ast}=R_{\sun}$ and $T_{\ast}=6000\,K$. In this regime the external radiation produces a hot surface dust layer which re-radiates $\sim1/2$ the absorbed light back into the disk, maintaining $T_{\rm mid}$; if the disk is optically thick to the incident and re-radiated emission, this gives an approximate $T_{\rm mid,\,\ast}\approx T_{\rm eff,\ast}^{1/4}/2^{1/4}$. Meanwhile, accretion produces an effective temperature 
\be
T_{\rm eff,\,acc} \approx {\Bigl[}\frac{3}{8\pi}\,\frac{\dot{M}\,\Omega^{2}}{\sigma_{\rm SB}} {\Bigr]}^{1/4}
\ee
In the optically thick limit, this is just related to $T_{\rm mid}$ by $T_{\rm mid,\,acc}^{4}\approx (3/4)\,(\tau_{R} + 2/3)\,T_{\rm eff,\,acc}^{4}$, where $\tau_{R}$ is the Rosseland-mean optical depth, $\tau_{R} = \kappa_{R}(T_{\rm mid})\,\Sigma$ (so $T_{\rm mid,\, acc}$ is determined implicitly). 

In the text, Figs.~\ref{fig:T.vs.r}-\ref{fig:sigma.vs.r} simply use the optically thick relations for $T_{\rm mid}$ and, when accretion is included, interpolate with $T_{\rm mid}^{4} \approx T_{\rm mid,\,\ast}^{4}+T_{\rm mid,\,acc}^{4}$. We approximate the opacities with the simple $\kappa_{R} \sim 5\,{\rm cm^{2}\,g^{-1}}$ at $T>160\,$K, and $\kappa_{R}\sim 2.4\times10^{-4}\,T^{2}\,{\rm cm^{2}\,g^{-1}\,K^{-2}}$ at lower temperatures (approximately what is obtained with a galactic gas-to-dust ratio; see \citealt{adams:1988.yso.spectra,bell:1994.protostellar.disk.model}). 

We then follow \citet{rafikov:2007.convect.cooling.grav.instab.planets} and estimate the cooling time as 
\be
t_{\rm cool} = \frac{\Sigma\,c_{s}^{2}}{\sigma_{\rm SB}\,T_{\rm mid}^{4}}\,f(\tau) 
\ee
where $c_{s} = \sqrt{k_{B}\,T_{\rm mid}/\mu}$ with $\mu\approx2.3$ appropriate for molecular, dusty gas. The function $f(\tau)$ is given by the interpolation between the convective and radiative terms at the photosphere 
%\begin{align}
%f(\tau)^{-1} &\approx f_{c}(\tau)^{-1} + f_{r}(\tau)^{-1} 
%\end{align}
%The convective term scales as 
\begin{align}
f(\tau) &= \chi\,\tau^{\eta} + \phi\,\tau^{-1} \\ 
\eta &= \frac{4\,\gamma^{-1}\,(\gamma-1)}{1+\alpha+\beta\,\gamma^{-1}\,(\gamma-1)}
\end{align}
here $\chi$ and $\phi$ are constants; from the detailed estimates therein $\phi\approx1$ and $\chi\approx0.19-0.31$, depending on the disk parameters, so we assume $\chi=0.31$ to be conservative (since this gives larger cooling times). This interpolates between the optically thick limits (dominated by $\chi\,\tau^{\eta}$) and optically thin cases ($\phi\,\tau^{-1}$, where the cooling flux becomes $\propto \tau\,\sigma_{\rm SB}\,T_{\rm mid}^{4}$). 
%The radiative term can be conveniently appro

The scaling index $\eta$ is determined from the temperature gradient to the photosphere for a disk in vertical hydrostatic equilibrium, under the assumption that over some (limited) local temperature range the opacity $\kappa$ can be approximated by $\kappa\approx \kappa_{0}\,P^{\alpha}\,T^{\beta}$, which is valid for our assumptions. We assume $\gamma=7/5$ in both this and the turbulent density fluctuation calculation in Figs.~\ref{fig:mass.vs.r}-\ref{fig:sigma.vs.r}.
Based on the scaling of opacities in \citet{semenov:2003.dust.opacities}, at $T<160\,$K, $\alpha=0$, $\beta\approx2$, so $\eta\approx7/11$; at $160 < T < 1500$\,K $\alpha=0$, $\beta\approx1/2-1$ and $\eta\approx7/8-7/9$ (we adopt $7/9$, but this makes no significant difference), and at $T>1500\,$K (when all grains sublimate and molecular opacity dominates) $\alpha\approx2/3$ and $\beta\approx7/3$, so $\eta\approx3/7$.

In the text, we restrict to these simple estimates because the quantities of interest are fairly uncertain. We can, however, examine a more detailed approximation here. First, we take the opacities $\kappa_{R}(T_{\rm mid})$ from the full tabulated values in \citet{semenov:2003.dust.opacities}. Second, a more accurate estimate of the midplane temperature is given by solving the implicit equation 
\be
T_{\rm mid}^{4} = \frac{3}{4}\,{\Bigl[}\tau_{V} + \frac{4}{3}+\frac{2}{3\,\tau_{V}} {\Bigr]}\,T_{\rm eff,\,acc}^{4} + {\Bigl[}1 + \frac{1}{\tau_{V}} {\Bigr]}\,T_{\rm eff,\,\ast}^{4}
\ee
where $\tau_{V}=\tau_{V}(T_{\rm mid})$ is the vertical optical depth from the midplane, $\tau_{V} = \kappa_{R}(T_{\rm mid})\,\Sigma/2$. This allows for an appropriate interpolation between the optically thick case and the case where disk is optically thin to its own re-radiation. Third, we can switch between the $f(\tau)$ above, appropriate for a convective disk, and $f(\tau) \approx \tau + \tau^{-1}$ appropriate for a purely radiative, convectively stable disk, when the disk falls below the (temperature-dependent) criteria for convective instability $\nabla_{0} \ge \nabla_{\rm ad}$ with $\nabla_{0}\equiv (1+\alpha)/(4-\beta)$ and $\nabla_{\rm ad}\equiv(\gamma-1)/\gamma$ \citep[see][]{lin.papaloizou.1980:mmsn,rafikov:2007.convect.cooling.grav.instab.planets}, where $\alpha$ and $\beta$ depend on $T$ (using the full explicit derivatives from the opacity tables). 

\begin{figure}
    \centering
    \plotonesize{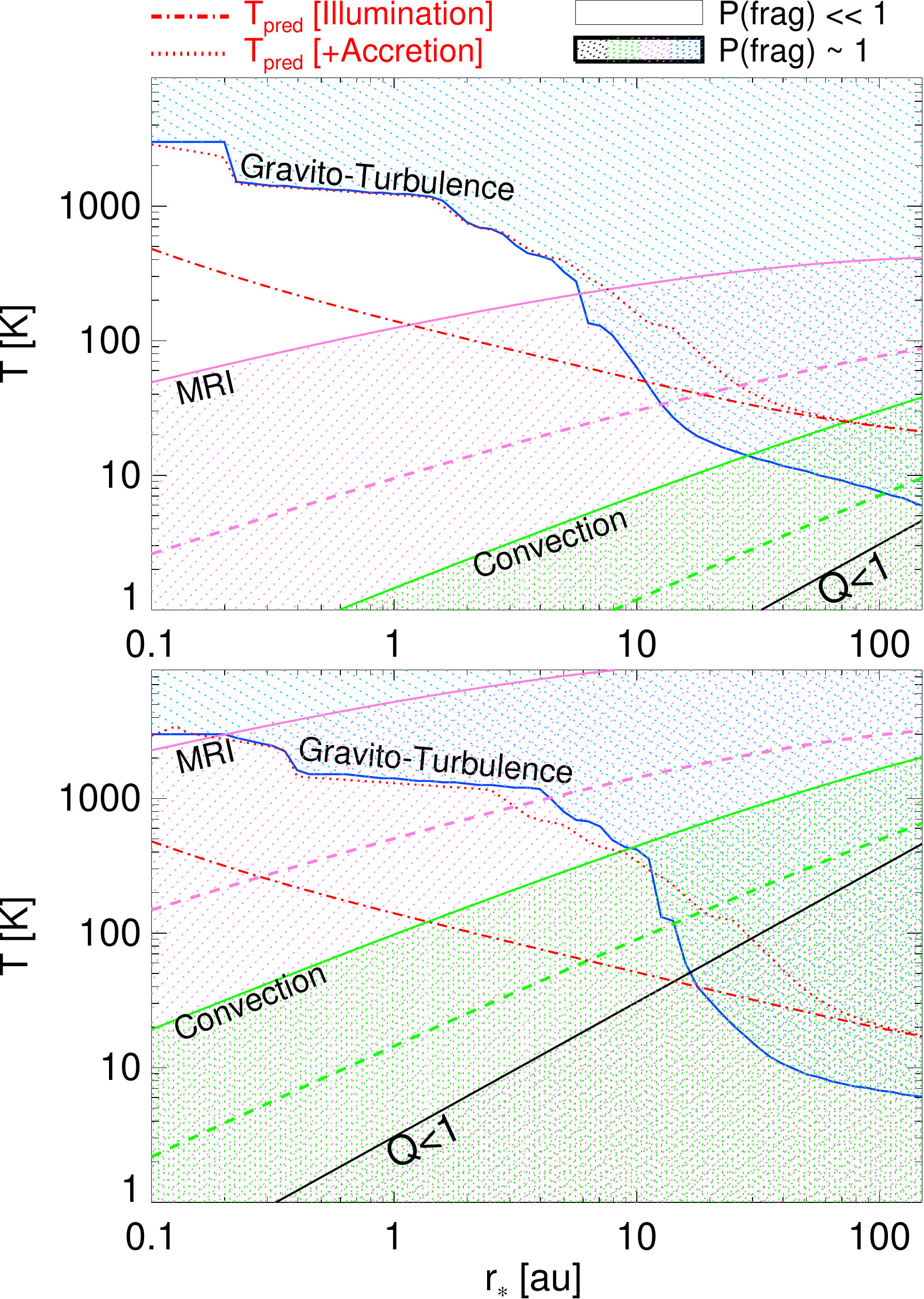}{1.01}
    \caption{Shaded regions show the temperature range in which there is statistical instability (order-unity probability of a collapse event), as Fig.~\ref{fig:T.vs.r}, but with a more detailed set of opacity tables and temperature calculation as described in \S~\ref{sec:appendix:tempcalc}.
    \label{fig:T.vs.r.kappa}}
\end{figure}

\begin{figure}
    \centering
    \plotonesize{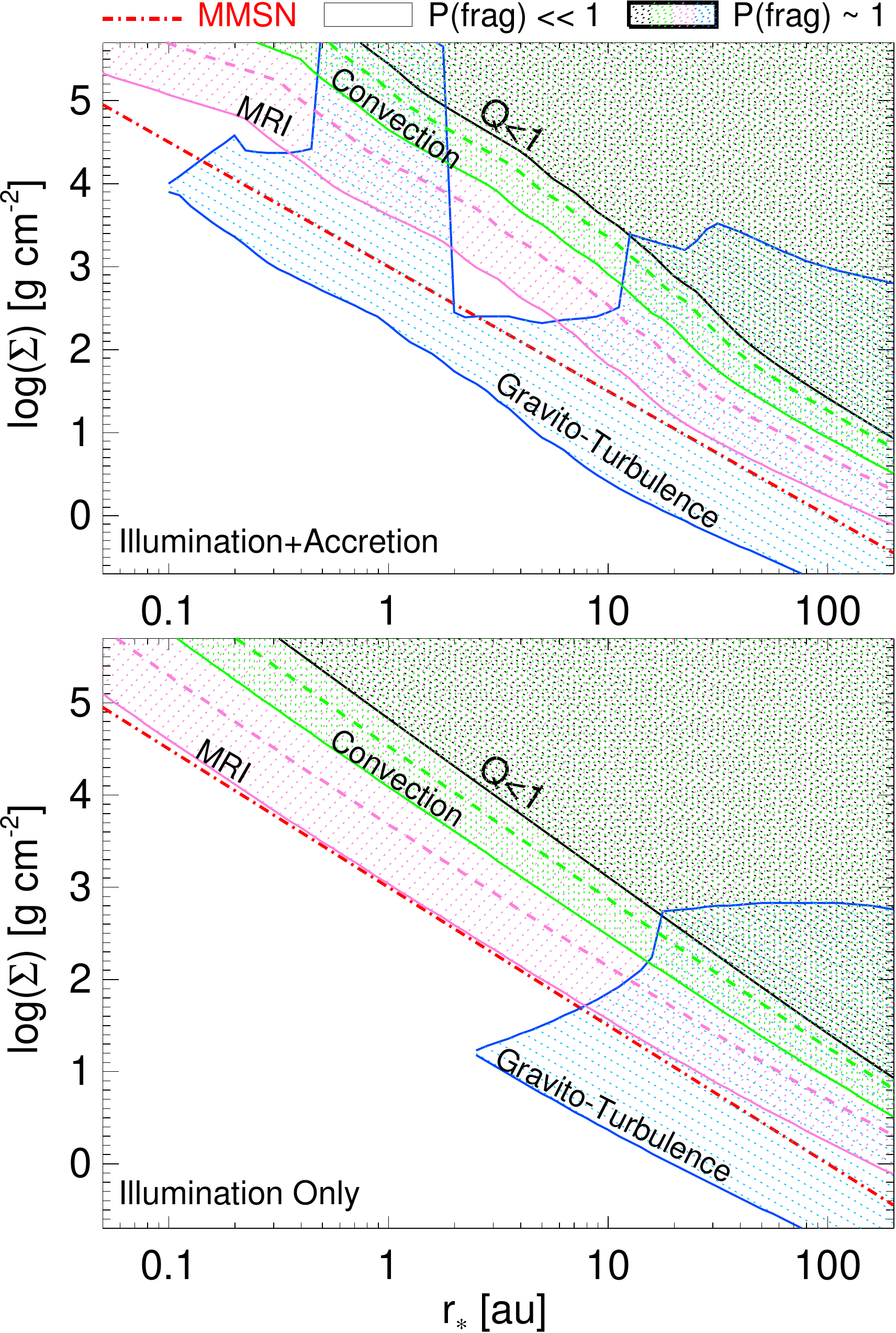}{1.01}
    \caption{Shaded regions show the surface density range in which there is statistical instability (order-unity probability of a collapse event), as Fig.~\ref{fig:sigma.vs.r}, but with a more detailed set of opacity tables and temperature calculation as described in \S~\ref{sec:appendix:tempcalc}.
    \label{fig:sigma.vs.r.kappa}}
\end{figure}

Figures~\ref{fig:mass.vs.r.kappa}-\ref{fig:sigma.vs.r.kappa} repeat our calculations from the main text with this more detailed temperature calculation. We find that the quantitative results in Figs.~\ref{fig:mass.vs.r}-\ref{fig:sigma.vs.r} are all changed at the factor $\lesssim2$ level and the qualitative conclusions are completely unchanged. The sense of the quantitative change tends to slightly expand the regions of parameter space where statistical instability and turbulence-promoted fragmentation can occur. The more detailed opacity calculation imprints some small features on the parameter space, the most significant of which is the elimination of predicted temperatures much larger than $\sim1500\,$K at small radii (because grains sublimate and cooling becomes optically thin), but the ``gravito-turbulent'' thresholds shift with the predicted temperatures so the statistical stability is essentially identical.

\clearpage
\vspace{-0.5cm}
\section{B.\ Overview of Additional Model Details}
\label{sec:appendix:model}

Here, we review the basic framework of the model developed in \paperone-\papertwo, specifically some key equations needed to reproduce the results in this paper. Readers interested in a full derivation and explanation of these equations should see \papertwo.

Consider, for simplicity, the isothermal (lognormal) case: if density fluctuations are lognormal, then the variable $\delta({\bf x})\equiv \ln{[\rho({\bf x})/\rhodisk]}+S/2$, 
where $\rho({\bf x})$ is the density at a point ${\bf x}$, 
$\rhodisk$ is the global mean density and $S$ is the variance in $\ln{\rho}$, 
is normally distributed according to the PDF:\footnote{The $+S/2$ term 
in $\delta$ is required so that the integral of $\rho\,P_{0}(\rho)$ correctly 
gives $\rhodisk$ with $\langle \delta\rangle = 0$.}
\be
P_{0}(\delta\,|\,S) = \frac{1}{\sqrt{2\pi\,S}}\,\exp{\left(-\frac{\delta^{2}}{2\,S} \right)}
\ee
More generally, we can evaluate the field $\delta({\bf x}\,|\,R)$, 
which is the $\delta({\bf x})$ field averaged around the point ${\bf x}$ with 
some window function of characteristic radius $R$; this is also normally distributed (see \papertwo, Appendix~F), with a variance at each scale $S(R)$ that is directly related to the density power spectrum. 

\paperone\ derives the ``first-crossing'' distribution for the general form of these fields. This corresponds to the mass and initial size spectrum of regions which are sufficiently dense so as to be self-gravitating averaged on the scale $R$ (specifically defined as the largest scale on which the region is self-gravitating, i.e.\ excluding bound sub-units already counted ``within'' the parent, although these can be counted separately if desired). This corresponds to the statistics of regions where $\delta({\bf x}\,|\,R) > B(R)$, where $B(R)$ (the ``barrier'') is some (scale-dependent) critical value. In \paperone\ we derive $S(R)$ and $B(R)$ from simple theoretical considerations for all scales in a galactic disk and/or molecular cloud. However, the derivation proceeds almost identically for a proto-planetary disk, following the most general form presented in \papertwo, which we outline here. 

It is well-established that the contribution to density variance from the velocity variance on a given scale goes as $S\approx\,\ln{(1+b^{2}\Mach^{2})}$, where $\Mach$ is the Mach number. For a given turbulent power spectrum, then, $S(R)$ is determined  
by summing the contribution from the velocity variance on all scales $R^{\prime}>R$:
\begin{align}
\label{eqn:S.R}
S(R,\,\rho) &= \int_{0}^{\infty} 
|\tilde{W}(k,\,R)|^{2}
\ln{{\Bigl [}1 + 
\frac{b^{2}\,v_{t}^{2}(k)}{c_{s}^{2}(\rho,\,k) + \kappa^{2}\,k^{-2}}
{\Bigr]}} 
{\rm d}\ln{k} 
\end{align}
where $W$ is the window function for the smoothing\footnote{{For convenience 
we take this to be a $k$-space tophat: $W=1$ for $k\le1/R$, $W=0$ otherwise. But we show in \paperone\ and \papertwo\ (Appendix~G) that this has little effect on our results. Similarly, we emphasize that whether the fluctuations are random-phase or correlated has little effect on our conclusions (as shown explicitly in \papertwo).}}, $v_{t}(k)$ is the turbulent velocity dispersion averaged on a scale $k$ (trivially related to the turbulent power spectrum), $c_{s}$ is the thermal sound speed (both $c_{s}$ and $S$ can depend locally on $\rho$ if the gas is not isothermal), and $b$ is the fraction of the turbulent velocity in compressive (longitudinal) motions (discussed below). Here $\kappa$ is the epicyclic frequency; since we are interested in Keplerian disks, we take $\kappa\approx\Omega$, the disk orbital frequency. Note that on large scales, angular momentum ($\kappa^{2}\,k^{-2}$) enters in a similar way to $c_{s}^{2}$ and suppresses fluctuations, which follows directly from the form of the dispersion relation for density perturbations \citep[e.g.][]{lin.shu:spiral.wave.dispersion,toomre:spiral.structure.review,lau:spiral.wave.dispersion.relations}; accounting for this is necessary to ensure mass conservation. 

Since we are interested in the formation of self-gravitating gas objects, we define $B(R)$ corresponding to the critical density averaged on a given scale, $\rhocrit(R)$, at which an overdensity will collapse. Given $\delta(R) \equiv \ln{[\rho(R)/\rhodisk]}+S/2$ defined above, then $B(R)$ follows from the dispersion relation for a density perturbation in a disk with self-gravity, turbulence, thermal and magnetic pressure, and angular momentum/shear \citep{vandervoort:1970.dispersion.relation,aoki:1979.gas.disk.instab.crit,elmegreen:1987.cloud.instabilities,romeo:1992.two.component.dispersion}: 
\be
B(R) = \ln{\left(\frac{\rhocrit(R)}{\rhodisk} \right)} + \frac{S(R)}{2}
\ee
where $\rhocrit$ is the critical density above which a region is self-gravitating. This is the (implicit) solution to 
\begin{align}
\label{eqn:rhocrit.appendix}
\frac{\rhocrit(R)}{\rhodisk} \equiv \frac{Q}{2\,\tilde{\kappa}}\,\left(1+\frac{h}{R} \right)
{\Bigl[} \frac{\sigma_{g}^{2}(R,\,\rhocrit)}{\sigma_{g}^{2}(h,\,\rhodisk)}\,\frac{h}{R}  + 
\tilde{\kappa}^{2}\,\frac{R}{h}{\Bigr]} 
\end{align}
where $\rhodisk$ is the mean midplane density of the disk, $h$ is the disk scale height, $\tilde{\kappa}\equiv\kappa/\Omega=1$ for a Keplerian disk, and 
$Q\equiv (\sigma_{g}[h,\,\rhodisk]\,\kappa)/(\pi\,G\,\Sigma_{\rm gas})$ is the Toomre $Q$ parameter. The ``total'' dispersion $\sigma_{g}$ is 
\be
\label{eqn:sigmagas}
\sigma_{g}^{2}(R,\,\rho) = c_{s}^{2}(\rho) + \langle v_{t}^{2}(R) \rangle  + v_{\rm A}^{2}(\rho,\,R)
\ee 

The map between scale $R$ and the total mass in the collapsing region is  
\be
\label{eqn:mass.radius}
M(R) \equiv 4\,\pi\,\rhocrit\,h^{3}\,
{\Bigl[}\frac{R^{2}}{2\,h^{2}} + {\Bigl(}1+\frac{R}{h}{\Bigr)}\,\exp{{\Bigl(}-\frac{R}{h}{\Bigr)}}-1 {\Bigr]}
\ee
It is easy to see that on small scales, these scalings reduce to the Jeans+Hill criteria 
for a combination of thermal, turbulent, and magnetic support, with $M=(4\pi/3)\,\rhocrit\,R^{3}$; on large scales it becomes a Toomre-like criterion 
with $M=\pi\Sigma_{\rm crit}\,R^{2}$. 

For any $B(R)$ and $S(R)$, \paperone\ shows that the instantaneous ``first-crossing'' mass function (i.e.\ instantaneous mass function of collapsing objects, uniquely defined to resolve the ``cloud-in-cloud'' problem) is 
\be
\frac{{\rm d}n}{{\rm d}M} = 
\frac{\rhocrit(M)}{M}\,\ffirst(M)\,{\Bigl |}\frac{{\rm d}S}{{\rm d}M} {\Bigr |}
\ee
where $\ffirst(S)$ is a function shown by \citet{zhang:2006.general.moving.barrier.solution} to be the solution of the Volterra integral equation:
\begin{align}
\label{eqn:ffirst}
\ffirst(S) = \tilde{g}_{1}(S) + \int_{0}^{S}\,{\rm d}S^{\prime}\,\ffirst(S^{\prime})\,\tilde{g}_{2}(S,\,S^{\prime})
\end{align}
with 
\begin{align}
\tilde{g}_{1}(S) &= {\Bigl [}2\,{\Bigl|}\frac{dB}{dS}{\Bigr|} +\frac{B(S)}{S}{\Bigr]}\,P_{0}(B(S)\,|\,S)\\
\tilde{g}_{2}(S,\,\Sprime) &= {\Bigl[}-\frac{B(S)-B(\Sprime)}{S-\Sprime} 
-2\,{\Bigl|}\frac{dB}{dS}{\Bigr|}{\Bigr]}\times\\
\nonumber& P_{0}[B(S)-B(\Sprime)\,|\,\Sprime-S]
\end{align}

\end{appendix}

\end{document}